\documentclass[aps,prc,onecolumn,superscriptaddress,showpacs,nofootinbib,floatfix]{revtex4}
\usepackage{dcolumn}
\usepackage{bm}
\usepackage[dvips]{graphicx}
\usepackage{float}
\usepackage{epsfig}
\usepackage{graphicx}
\usepackage{longtable} 
\usepackage{rotating}
\usepackage{amsmath}
\usepackage{amsfonts}
\usepackage{amssymb}

\newcommand{\sqrts}{\sqrt{s}}
\newcommand{\sqrtsnn}{\sqrt{s_{_{\ensuremath{\it{NN}}}}}}
\def\mean#1{\ensuremath{\left<#1\right>}}
\def\ttt#1{\texttt{\small #1}}
\newcommand{\dd}{d}
\def\X{{\rm X}}
\newcommand{\ndf}{{\rm ndf}}
\def\cO#1{{{\cal{O}}}\left(#1\right)}

\begin{document}

\title{Single inclusive pion $p_{_T}$-spectra in proton-proton collisions\\[0.3cm] at $\sqrt{s}$~=~22.4~GeV: data versus perturbative QCD calculations}

\author{Fran\c{c}ois~Arleo}

\affiliation{
LAPTH\footnote{Laboratoire d'Annecy-le-Vieux de Physique Th\'eorique, UMR5108}, 
Universit\'e de Savoie, CNRS, BP 110, 74941 Annecy-le-Vieux cedex, France}

\author{David~d'Enterria}

\affiliation{
CERN, PH Dept., 1211 Geneva 23, Switzerland}

\begin{abstract}
\noindent
We compare the inclusive transverse momentum spectra of single pions above 
$p_{_T}$~=~3~GeV$/c$ measured in proton-proton ($p$--$p$) collisions at 
$\sqrt{s}$~=~21.7 --~23.8~GeV, with next-to-leading order (NLO) perturbative QCD (pQCD)
predictions using recent parametrizations of the parton densities and parton-to-pion 
fragmentation functions. Although the dependence on the theoretical scales is large, 
the calculations can reproduce the experimental results both in magnitude and shape.
Based on the existing data and on a pQCD $\sqrt{s}$-rescaling of the measured spectra, 
we provide a practical parametrization of the baseline $p$--$p$ pion transverse momentum spectrum 
to be compared to nucleus-nucleus collisions data at $\sqrtsnn$~=~22.4~GeV.
\end{abstract}


\maketitle


\section{Introduction}

The study of hadron production at large transverse momenta 
($p_{_T}\gg\Lambda_{\rm QCD}\approx$~0.2~GeV)
in  hadronic interactions is a valuable testing ground of the perturbative regime 
of Quantum Chromodynamics (pQCD), providing information on both the parton 
distribution functions (PDFs) in the proton, and the parton-to-hadron fragmentation 
functions (FFs)~\cite{geist90}. In the last years, a renovated interest in high-$p_{_T}$ 
hadron production has been driven mainly by studies of ``jet quenching'' phenomena
in high-energy nucleus-nucleus (A-A) collisions~\cite{dde_jetquench} as well as 
of the proton spin structure in polarized $p$--$p$ collisions~\cite{Bunce:2000uv,Jager:2002xm}. 
In A-A collisions, the observed large suppression of high-$p_{_T}$ hadron yields compared 
to (appropriately scaled) $p$--$p$ cross sections~\cite{Adler:2003qi,Adams:2003kv}
-- attributed to parton energy loss due to medium-induced gluon radiation~\cite{BDMPS,Gyulassy:2003mc} -- 
provides valuable information on the transport properties of hot and dense QCD matter~\cite{dde_jetquench}. 
The energy density at which such ``jet quenching'' phenomena sets in in A-A collisions 
can signal the possible transition from a hadronic to a deconfined quark-gluon system. 
Whereas unambiguous signals of high-$p_{_T}$ hadron suppression have been found at 
the Relativistic Heavy-Ion Collider (RHIC) in central Au-Au collisions at $\sqrtsnn$~=~200~GeV~\cite{Adler:2003qi,Adams:2003kv} 
and 62.4~GeV~\cite{Isobe:2005mh,Abelev:2007ra}, one cannot draw any firm conclusion yet 
at the Super Proton Synchrotron (SPS) energies ($\sqrtsnn$~=~17.3~GeV)~\cite{Blume:2006va} due to the lack of a valid 
(experimental and/or theoretical) proton-proton reference~\cite{d'Enterria:2004ig}. Recently, 
the PHENIX collaboration has presented results on high-$p_{_T}$ neutral pion production 
in Cu-Cu collisions at energies, $\sqrtsnn$~=~22.4~GeV, close to the SPS range~\cite{Adare:2008cx}.
We present here an experimental and theoretical study of the pion $p_{_T}$-spectrum 
in $p$--$p$ collisions required in order to determine the associated nuclear ``suppression factor'', 
$R_{AA}(p_{_T})\propto(dN_{AA}/dp_{T})/(dN_{pp}/dp_{_T})$ in A-A collisions at this 
center-of-mass (c.m.) energy.\\

In Section~\ref{sec:exp}, we compile and examine all existing experimental spectra for 
$\pi^0$~\cite{carey76,bonesi89,demarzo87,donaldson78,lloydowen80,busser76,eggert75,alper75}
and $\pi^\pm$~\cite{alper75,adamus88} at c.m. energies in the range 
$\sqrt{s}$~=~21.7 --~23.8~GeV. We notice that most of the data appear to be consistent with each 
other within uncertainties, despite some spread. In Section~\ref{sec:pqcd}, we compare these data to pQCD calculations at 
Next-to-Leading Order (NLO) accuracy, as implemented in the Monte Carlo programme INCNLO~\cite{incnlo,Aurenche:1999nz}. 
We discuss in some detail the improvements in the model predictions thanks to the
use of recent FFs~\cite{albino:2008af}. For a choice of renormalization-factorization scales in the low side
($\mu/p_{_T}=1/3-1/2$), the calculations can reproduce the experimental results both in magnitude and shape 
within 
the uncertainties associated with the limited knowledge of the parton-to-pion fragmentation functions (FFs) 
and parton distribution functions (PDFs) in this kinematic range.
Finally, a practical parametrization of the $p$--$p$ pion transverse 
momentum spectrum at $\sqrt{s}$~=~22.4~GeV is provided in Section~\ref{sec:fit} for use as denominator in the determination of the 
corresponding nuclear modification factor in A-A collisions
in the low range of energies accessible at the RHIC collider.


\section{Inclusive pion spectra in $p$--$p$ collisions at $\sqrt{s}\approx$~22.4~GeV: Experimental measurements}
\label{sec:exp}

Table~\ref{tab:compilation} compiles the 13 measurements found in the literature for
neutral~\cite{carey76,bonesi89,demarzo87,donaldson78,lloydowen80,busser76,eggert75,alper75}
and charged~\cite{alper75,adamus88} pion production  at c.m. energies 
around $\sqrt{s}$~=~22.4~GeV at mid-rapidity ($y$~=~0, corresponding
to laboratory angles $\theta_{\rm lab}\approx$~1.~rad in fixed-target kinematics). 
The data were measured in the 70's at the CERN Intersecting Storage Rings (ISR) collider 
as well as in the 80's in various CERN and Fermilab (FNAL) fixed-target experiments. 
 The corresponding data points (adding to a total 
of $\sim$220) have been obtained from the Durham database~\cite{durham}.
Assuming isospin symmetry, the $\pi^0$ yield is 
the same 
as the $(\pi^++\pi^-)/2$ yield, and thus we can use both data sets to get a combined pion 
reference spectrum. The last column of Table~\ref{tab:compilation} collects the propagated
experimental uncertainties of the measurements as reported in the original publications. 
Two types of errors are often quoted: (i) those related to energy scale ($p_{_T}$) uncertainties, 
and (ii) additional systematic and/or absolute normalization (usually luminosity) errors. 
The $p_{_T}$-scale uncertainties have been transformed into an associated absolute cross 
section uncertainty assuming a local power-law distribution with exponent $n\approx$~10. 
We have conservatively added all quoted uncertainties in quadrature with the point-to-point
errors. We note that at variance with the $\pi^0$ spectra measured at 
$\sqrt{s}\approx$~63~GeV~\cite{dde_hq04}, there is no need to account for 
possible direct-$\gamma$ contaminations in the oldest ``non-resolved''  pion spectra since, 
at the lower c.m. energies considered here, the prompt photon contributions start to be 
significant only {\it above} the momentum range ($p_{_T}\gtrsim6$~GeV$/c$) actually 
reached in the experiments.\\

\setlength{\tabcolsep}{1pt}
\begin{table}[htpb]
\centering 
\begin{tabular}{c|c|c|c|c|c|c|c}\hline\hline 
\hspace{2mm} Reaction \hspace{2mm} & \hspace{3mm} $\sqrt{s}$ \hspace{3mm} & \hspace{3mm} $p_{\rm lab}/c$ \hspace{3mm} & \hspace{2mm} Collab./Exp. \hspace{2mm} & \hspace{2mm} Ref. \hspace{2mm} & \hspace{2mm} $p_{_T}$ range \hspace{2mm}  & \hspace{0mm} \# data \hspace{0mm} & Syst.\\
 & (GeV) & (GeV) &  &  & (GeV$/c$) & points & uncertainties \\\hline
$pp \rightarrow \pi^{0}$  X & 21.7  &  250. & FNAL E-063 & \cite{carey76} & 0.7 -- 2.4 & 29 & 30\% \\
$pp \rightarrow \pi^{-}$  X & 21.7  &  250. & EHS-NA22 & \cite{adamus88} & 0.1 -- 2.2 & 45 & -- \\
$pp \rightarrow \pi^{0}$  X & 22.8  &  275. & FNAL E-063 & \cite{carey76} & 0.4 -- 3.8 & 16 & 30\%  \\
$pp \rightarrow \pi^{0}$  X & 23.0  &  280. & CERN-WA70 & \cite{bonesi89} & 4.1 -- 6.7 & 8 & 16--30\% \\
$pp \rightarrow \pi^{\pm}$  X & 23.0  &  - & Brit.-Scand. & \cite{alper75} & 0.2 -- 3.0 & 17 & 15\%\\
$pp \rightarrow \pi^{0}$  X & 23.0  &  284. & FNAL E-063 & \cite{carey76} & 0.4 -- 4.5 & 14 & 30\% \\
$pp \rightarrow \pi^{0}$  X & 23.3  &  - &  R-107 & \cite{lloydowen80} & 1.0 -- 3.0 & 21 & 35\%\\ 
$pp \rightarrow \pi^{0}$  X & 23.5  &  - &  CCRS & \cite{busser76} & 2.5 -- 4.0 & 17 & 26\% \\
$pp \rightarrow \pi^{0}$  X & 23.6  &  - &  ACHM & \cite{eggert75} & 0.7 -- 4.5 & 19 & 35\% \\ 
$pp \rightarrow \pi^{0}$  X & 23.8  &  300. & FNAL E-063 & \cite{carey76} & 0.4 -- 3.7 & 12 & 30\% \\
$pp \rightarrow \pi^{0}$  X & 23.8  &  300. & CERN-NA24 & \cite{demarzo87} & 1.25 -- 6.0 & 9 & 15\% \\
$pp \rightarrow \pi^{0}$  X & 23.8  &  300. & FNAL-E-268 & \cite{donaldson78} & 1.3 -- 4.2 & 10 & 5\%\\ 
\hline 
\hline 
\end{tabular}
\caption{Compilation of inclusive pion production data in $p$--$p$ collisions around $\sqrt{s}$~=~22.4~GeV and midrapidity: 
collision, center-of-mass energy, $p_{lab}$ (for fixed-target experiments), collaboration/experiment name, 
bibliographical reference, measured $p_{_T}$ range, total number of data points, and associated systematic uncertainties 
in the measured cross sections.}
\label{tab:compilation} 
\end{table}

\begin{figure}[htpb]
\begin{center}
\includegraphics[height=10.0cm]{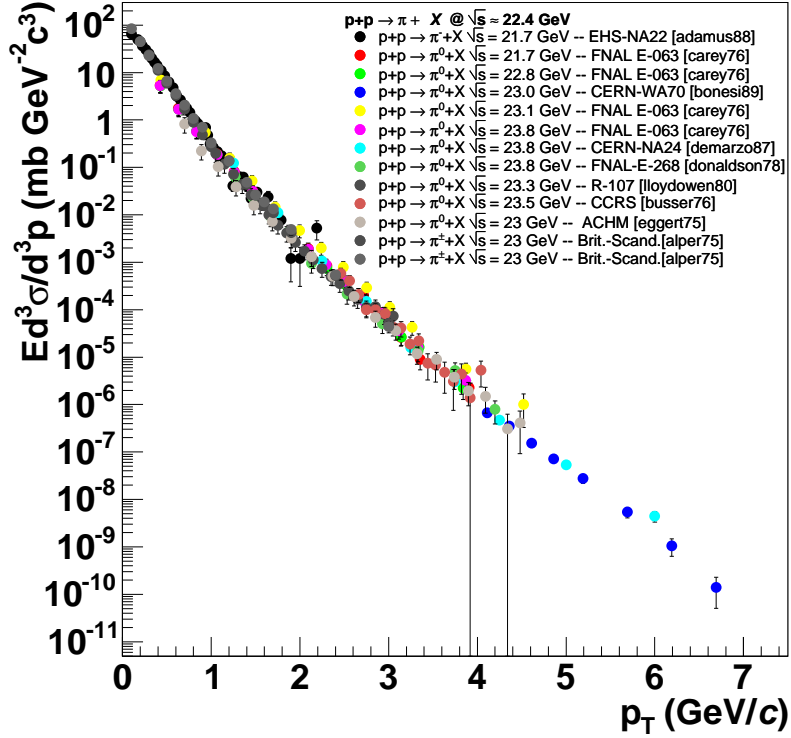}
\end{center}
\caption{Compilation of all pion transverse spectra measured in $p$--$p$ collisions in the
range $\sqrt{s}$~=~21.7 -- 23.8~GeV (see Table~\ref{tab:compilation} for details).}
\label{fig:all_spectra}
\end{figure}

Figure~\ref{fig:all_spectra} shows all the measured pion $p_{_T}$ spectra. The full range of cross sections
covers more than 12 decades. Unlike
with what was observed at $\sqrt{s}\approx$~63~GeV~\cite{dde_hq04}, 
the data taken by the various experiments appear in general quite compatible with each other both in shape 
and absolute cross sections,
within the experimental uncertainties and within the differences expected (at high-$p_{_T}$) 
due to the slightly dissimilar c.m. energies of the various measurements (see section~\ref{sec:rescaling}).
The spectra are characterized by an exponential distribution (with inverse slope $\sim$150~MeV) at low-$p_{_T}$ 
($p_{_T}\lesssim 1$~GeV$/c$), followed by a power-law with exponent $\sim 10$, and then a drop at the
highest $p_{_T}$'s when running out of phase-space for particle production, approaching the kinematical limit 
($p_{_T}^{\rm max} = \sqrt{s}/2$~=~11--12~GeV$/c$ at midrapidity).

\section{Inclusive pion spectra in $p$--$p$ collisions at $\sqrt{s}\approx$~22.4~GeV: NLO pQCD calculations}
\label{sec:pqcd}

The inclusive cross section for the production of a single pion, differential in transverse momentum 
$p_{_T}$ and rapidity $y$, takes the following form at next-to-leading order (NLO)~\cite{Aurenche:1999nz}:

\begin{eqnarray}
{\dd\sigma\over \dd{\bf p_{_T}}\dd{y}} &=& \sum_{i,j,k=q,g} \int \dd{x_1}\ \dd{x_2}\ 
F_{i/p}(x_1, \mu_{F})\ F_{j/p}(x_2,\mu_{F})\ {\dd{z} \over z^2}\ D_k^\pi(z,\mu_{\,\ensuremath{\it{ff}}}) 
\nonumber \\
&&{}\times \left [\left ( {\alpha_s (\mu_{R} ) \over 2 \pi} \right )^2
{\dd\widehat{\sigma}_{ij,k} \over \dd{\bf p_{_T}} \dd{y}}
+ \left ( {\alpha_s(\mu_{R}) \over 2 \pi} \right )^3 K_{ij,k}(\mu_{R} , \mu_{F}, \mu_{\,\ensuremath{\it{ff}}})
\right ]. 
\label{eq:dsigma_pQCD}
\end{eqnarray}

$F_{i/p}$ are the parton distribution functions (PDFs) of the incoming protons $p$,
$D_k^\pi(z, \mu_{\,\ensuremath{\it{ff}}})$ are the parton-to-pion fragmentation functions (FFs) describing 
the transition of the parton $k$ into a pion, 
and 
$\dd\widehat{\sigma}_{ij,k}/\dd{\bf p_{_T}} \dd{y}$ is the Born cross section of the 
subprocess $i + j \to k + \X$, and $K_{ij,k}$ is the corresponding higher-order term (the full kinematic dependence is omitted for clarity). 
In this paper, we use the INCNLO program~\cite{incnlo} to compute the cross sections, 
supplemented with various PDFs and FFs sets (see below).
The truncation of the perturbative series at next-to-leading order accuracy in
$\alpha_s$, introduces an artificial dependence, with magnitude $\cO{\alpha_s^3}$, of the cross section on initial- and final-state 
factorization scales, $\mu_{F}$ and $\mu_{\,\ensuremath{\it{ff}}}$, as well as on the renormalization scale $\mu_{R}$. 
The choice of scales is to a large extent arbitrary. One often uses as ``standard''
choice the hard scale of the process, e.g. $\mu_{R}=\mu_{F}=\mu_{\,\ensuremath{\it{ff}}}~=p_{_T}$. 
A more theoretically sound solution is given by using the Principle of Minimum 
Sensitivity (PMS)~\cite{PMS}. Phenomenological comparisons of pQCD results 
at various orders (LO, NLO, NNLO) among each other and against various 
experimental data sets (for charm and beauty, top, $Z$, $W$ bosons, ...) indicate that 
choosing a relatively low range of scales  $\mu/p_{_T}=1/3$--$1/2$ provides 
effectively a reduced sensitivity to higher-order effects~\cite{ageiser07}. 
We thus use $\mu_{R}=\mu_{F}=\mu_{\,\ensuremath{\it{ff}}}=p_{_T}/\kappa$, with variation between 
$\kappa=2$--$3$.  At small $p_{_T}$ and for the scale $p_{_T}/3$, the factorization 
scale approaches the starting scale $Q_0$ of the PDF evolution, where the parton 
densities are not constrained by data. To avoid this problem, 
we only compute the pion spectra above\footnote{Whenever it becomes smaller than the minimum 
$Q_0$ allowed by the PDF or FF parametrization, the hard scale $Q$ is frozen at $Q_0$.} $p_{_T}=3$~GeV$/c$. \\

\begin{figure}[htbp]
\begin{center}
\includegraphics[width=7.cm,height=6.cm]{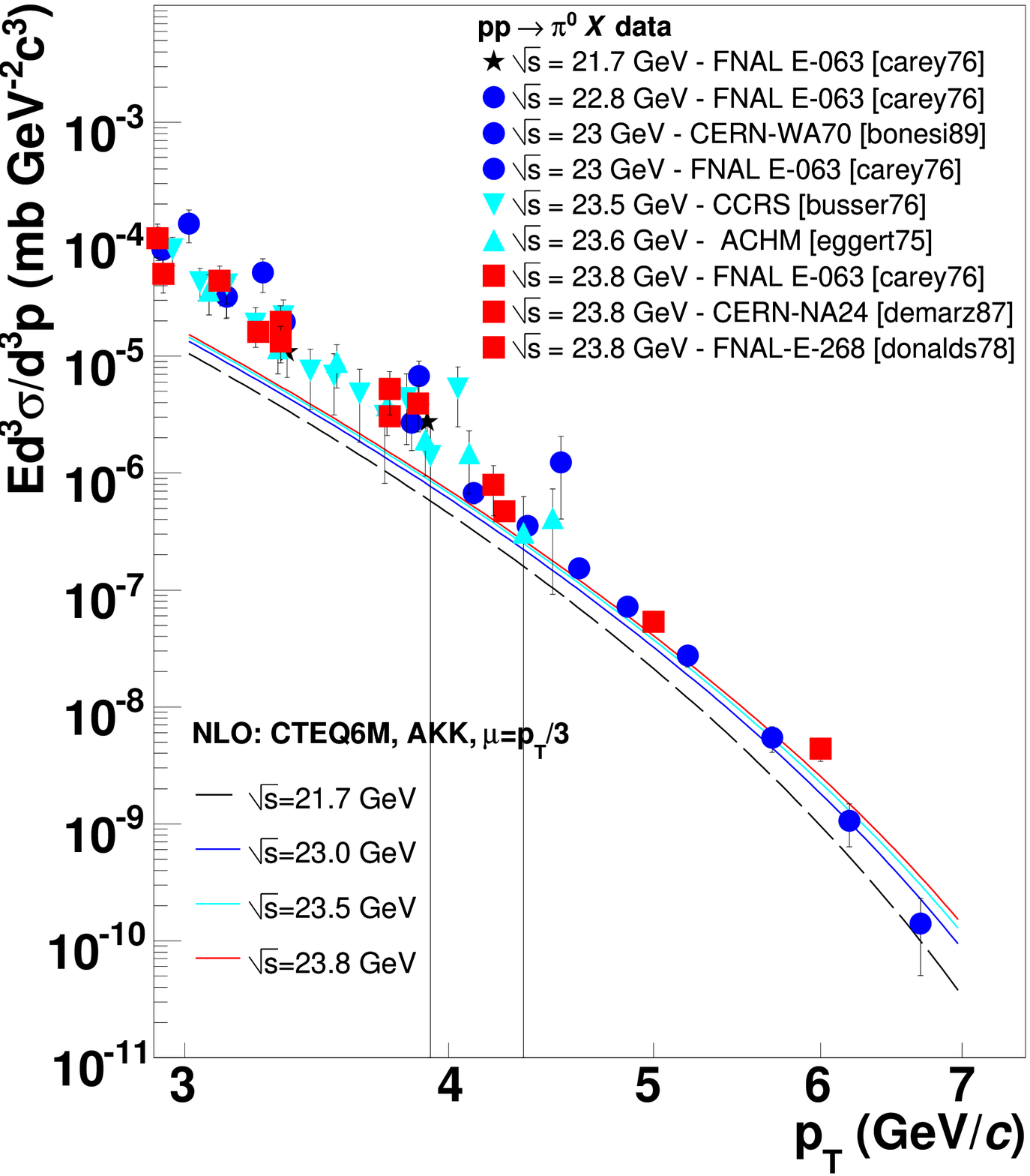}
\includegraphics[width=7.cm,height=6.cm]{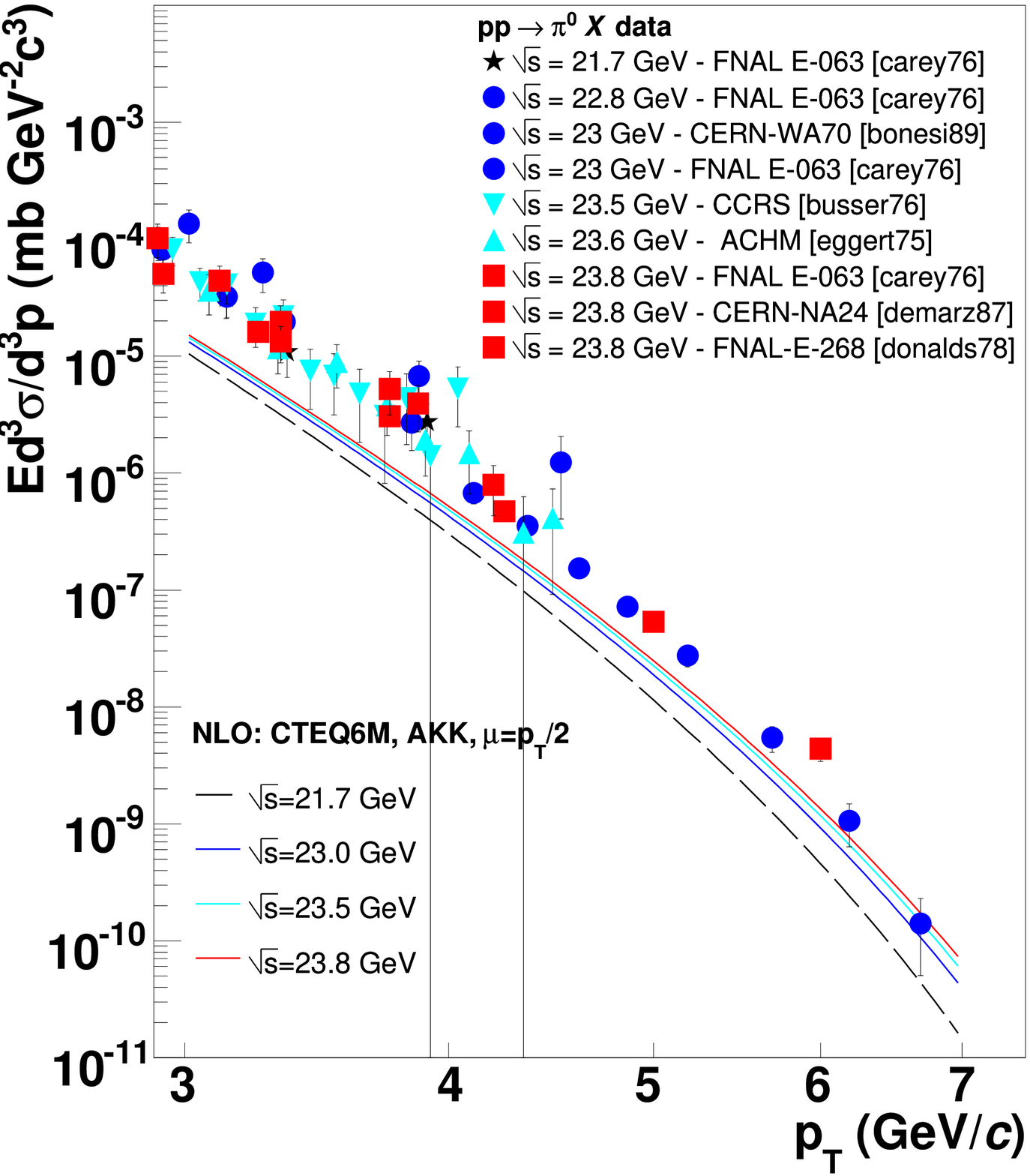}
\includegraphics[width=7.cm,height=6.cm]{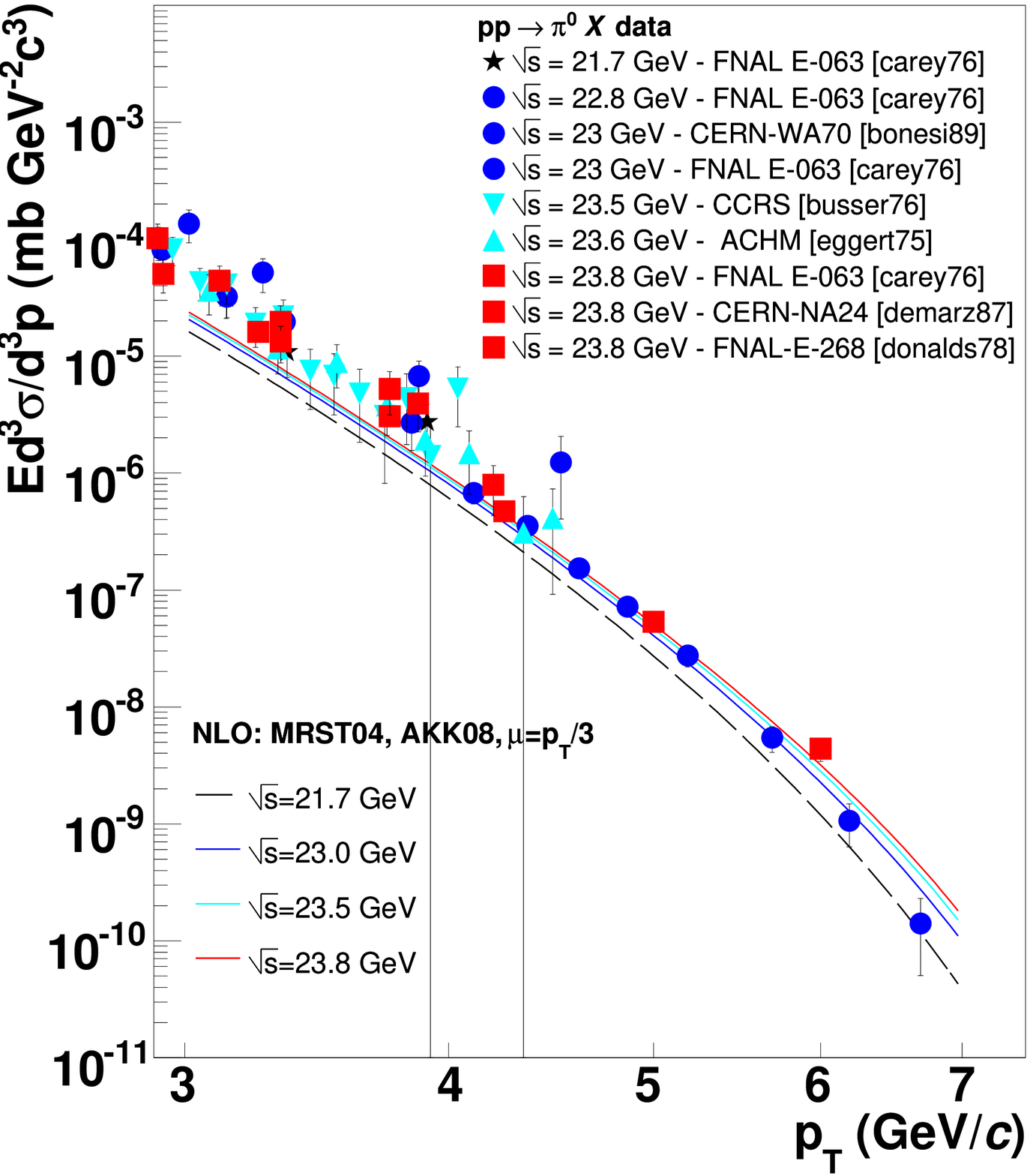}
\includegraphics[width=7.cm,height=6.cm]{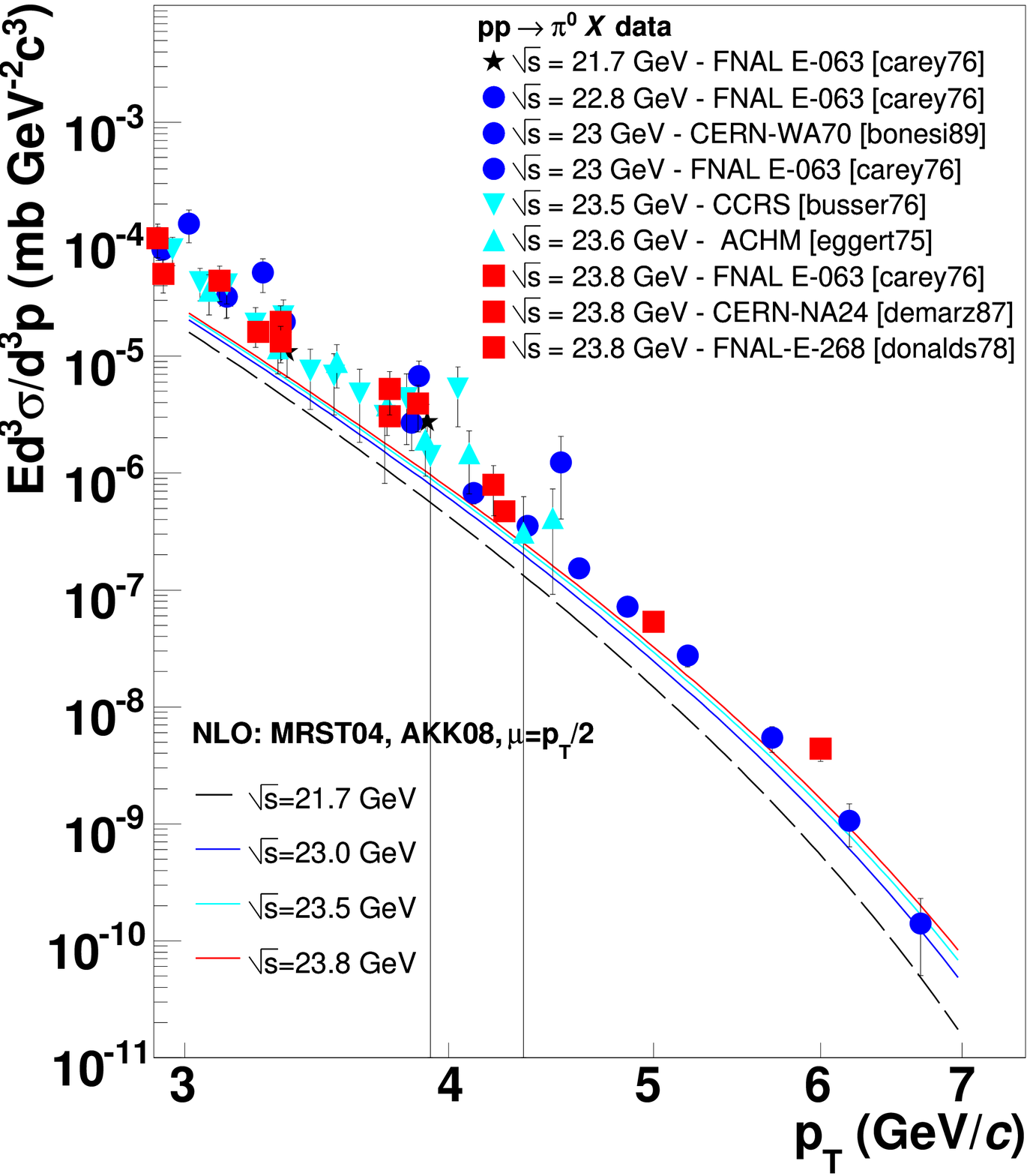}
\includegraphics[width=7.cm,height=6.cm]{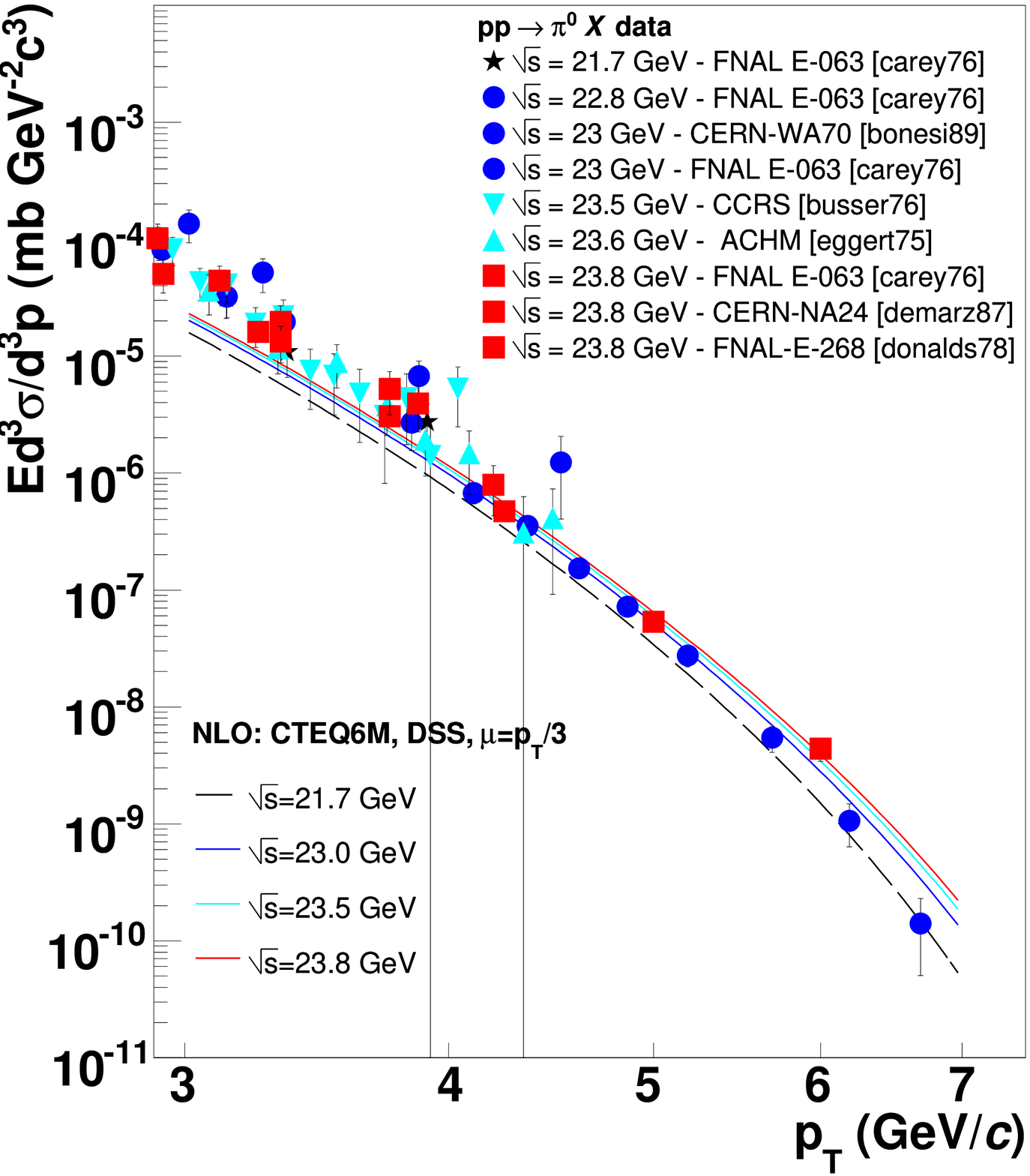}
\includegraphics[width=7.cm,height=6.cm]{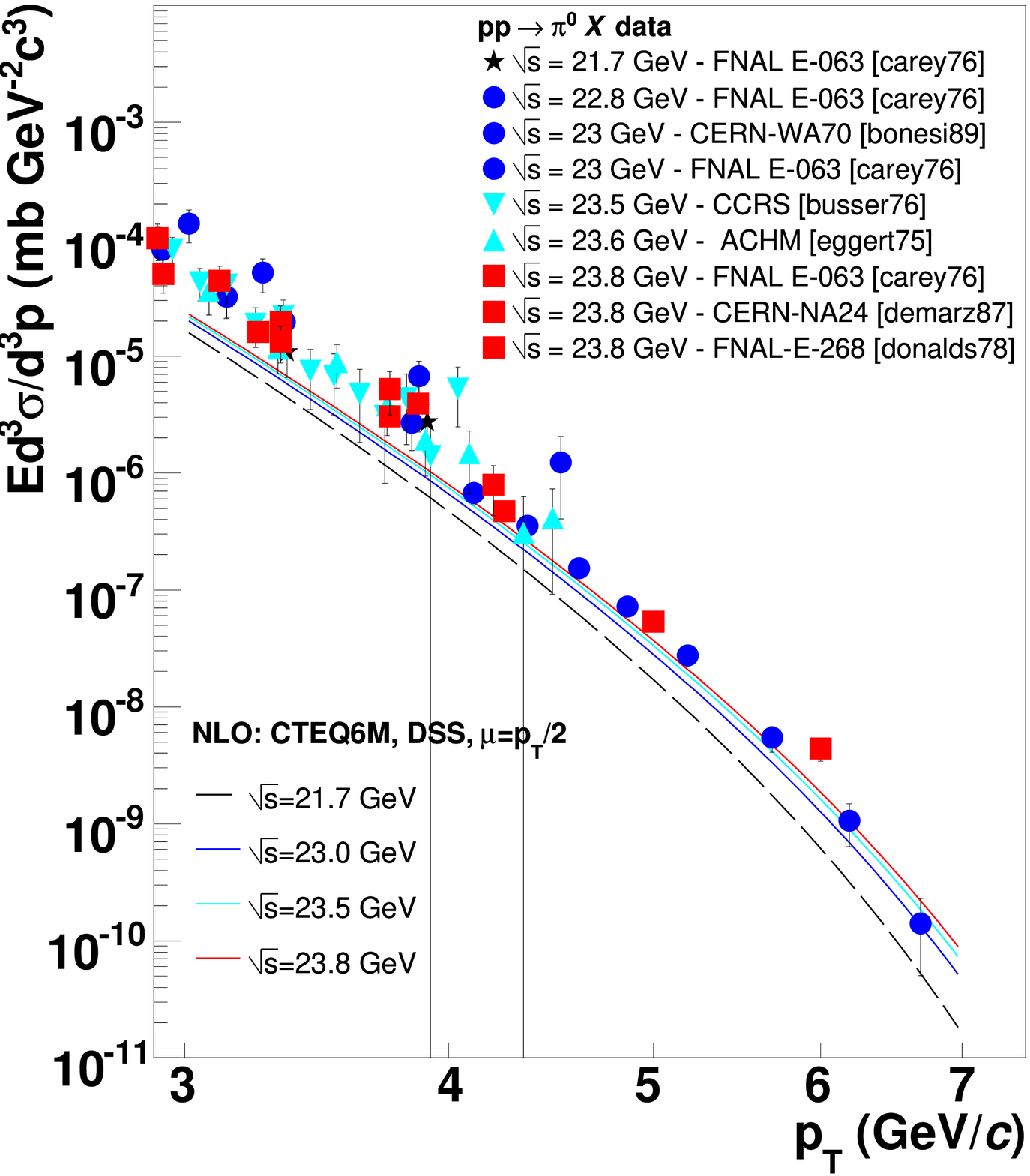}
\includegraphics[width=7.cm,height=6.cm]{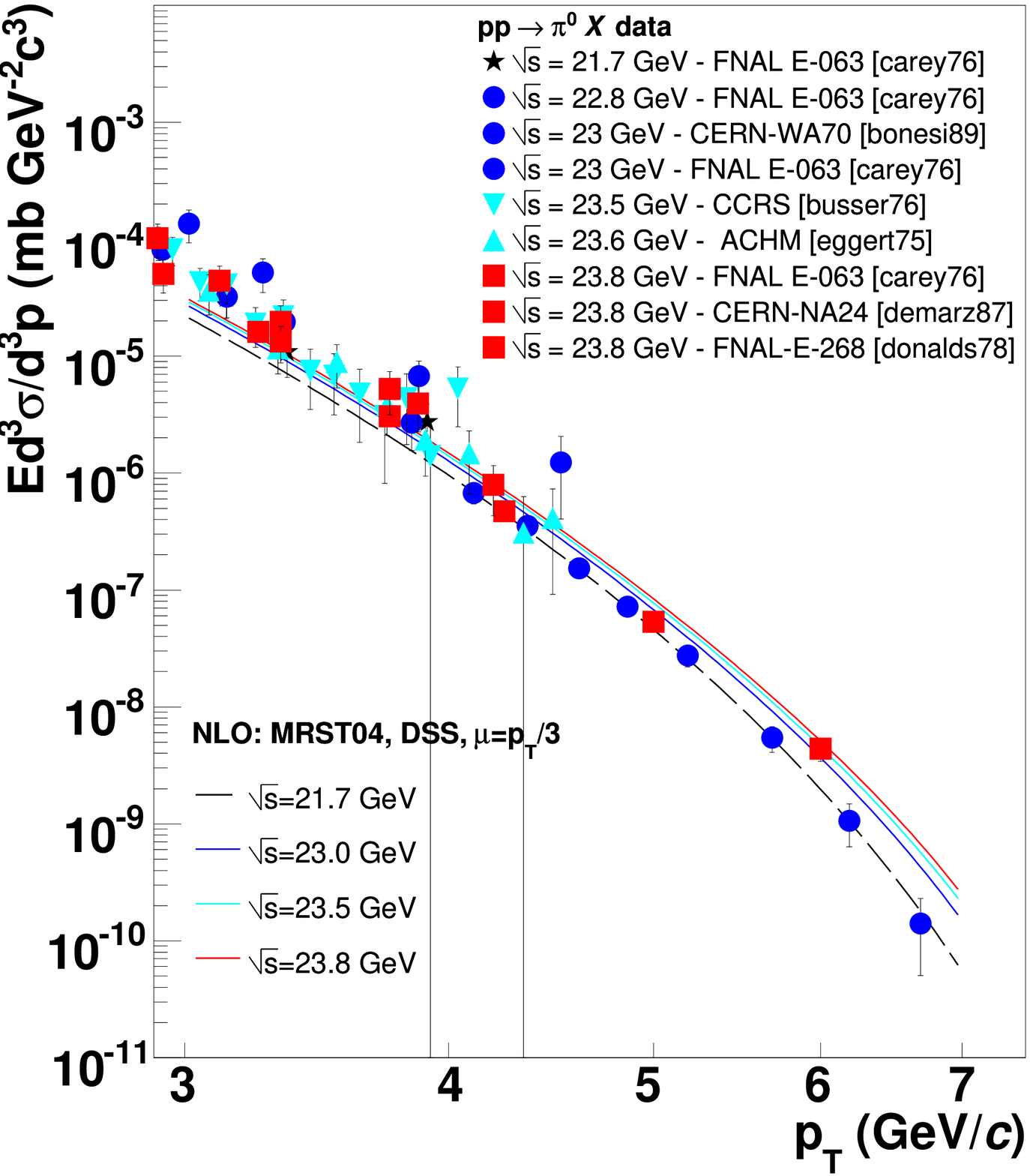}
\includegraphics[width=7.cm,height=6.cm]{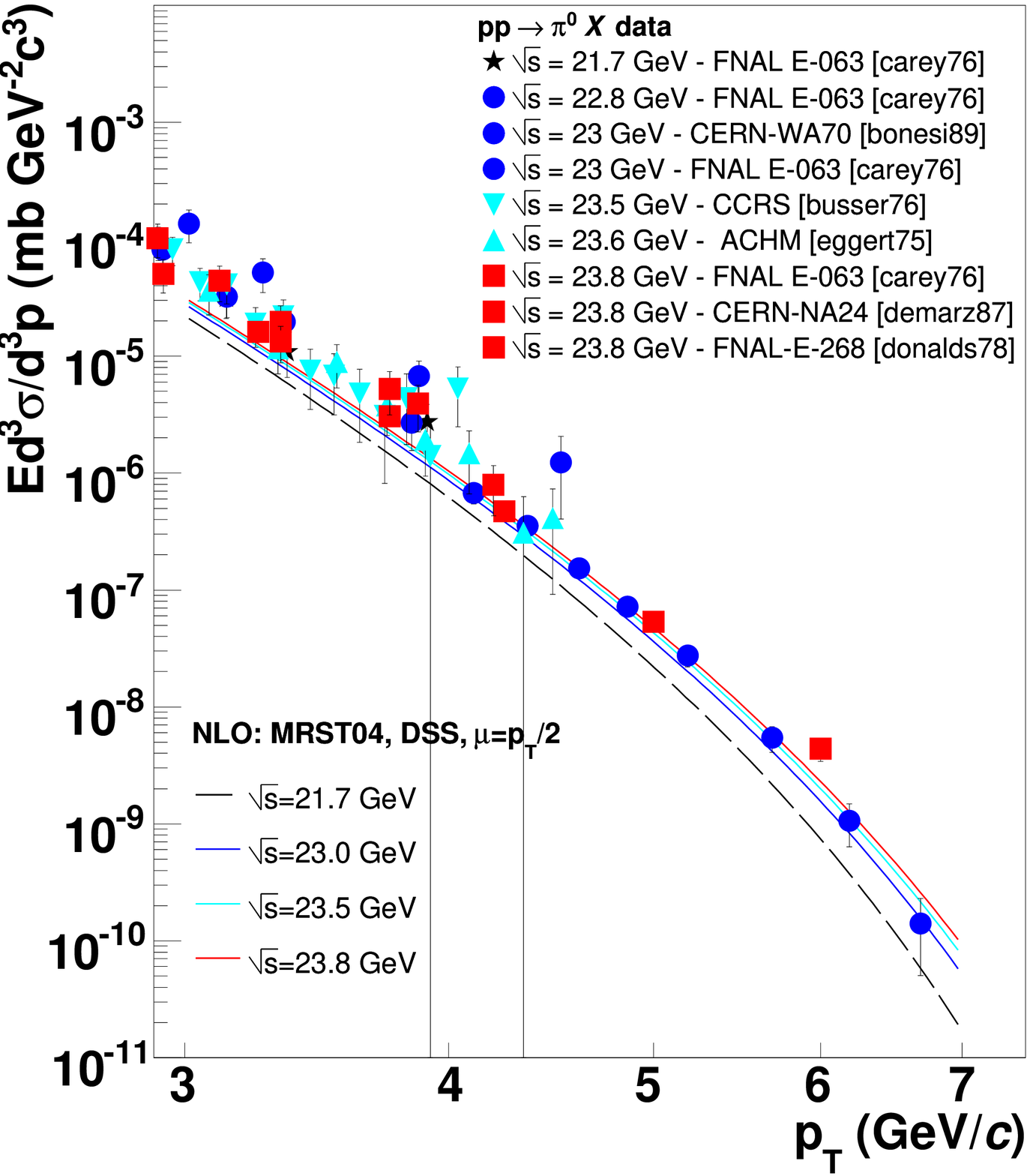}
\end{center}
\vspace{-0.5cm}
\caption{Comparison of pion transverse spectra measured in $p$--$p$ collisions at $\sqrt{s}\approx$~21.7--23.8~GeV 
to NLO pQCD predictions. The left (right) plots are for theoretical scales $\mu~=~p_{_T}/3$ ($p_{_T}/2$). 
Two sets of PDFs (MRST04 and CTEQ6.1M) and three FFs (AKK05, AKK08, DSS, from top to bottom) are used.}
\label{fig:data_pQCD}
\end{figure}

\begin{figure}[htbp] 
\begin{center}
\includegraphics[width=6.0cm,height=4.3cm, bb=0 0 360 304,clip]{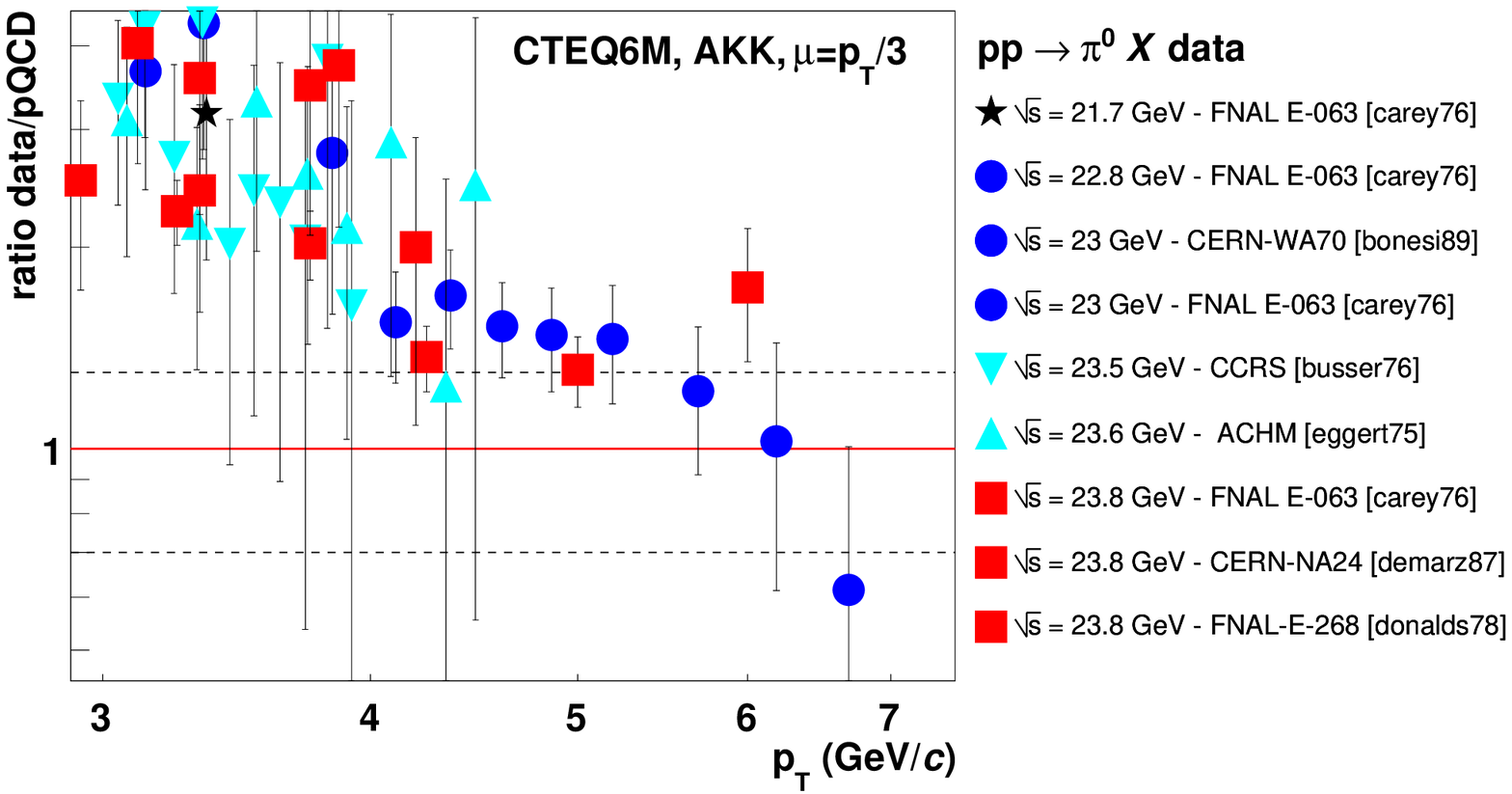}
\includegraphics[width=8.9cm,height=4.3cm]{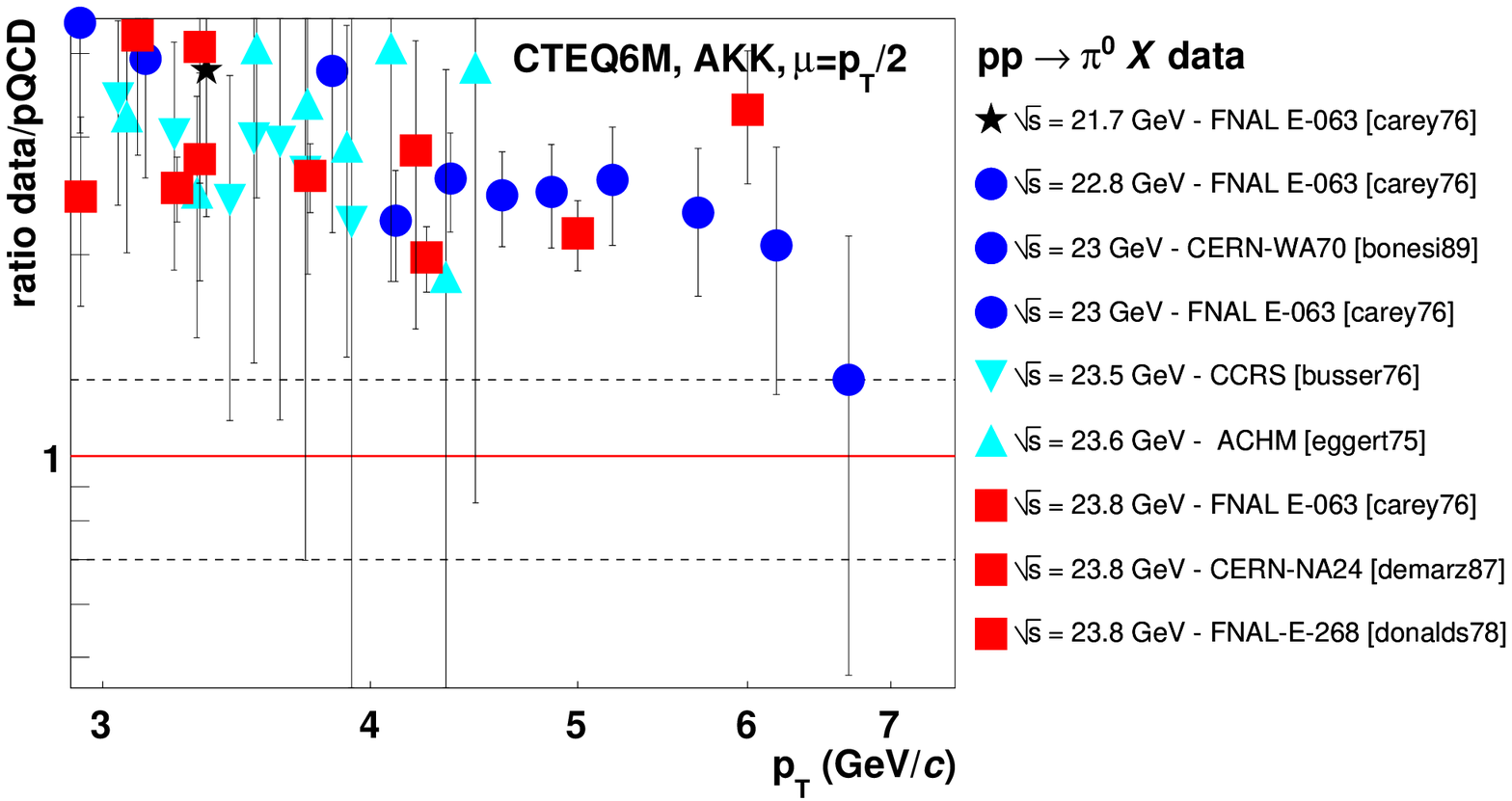}\\
\includegraphics[width=6.0cm,height=4.3cm,bb=0 0 360 304,clip]{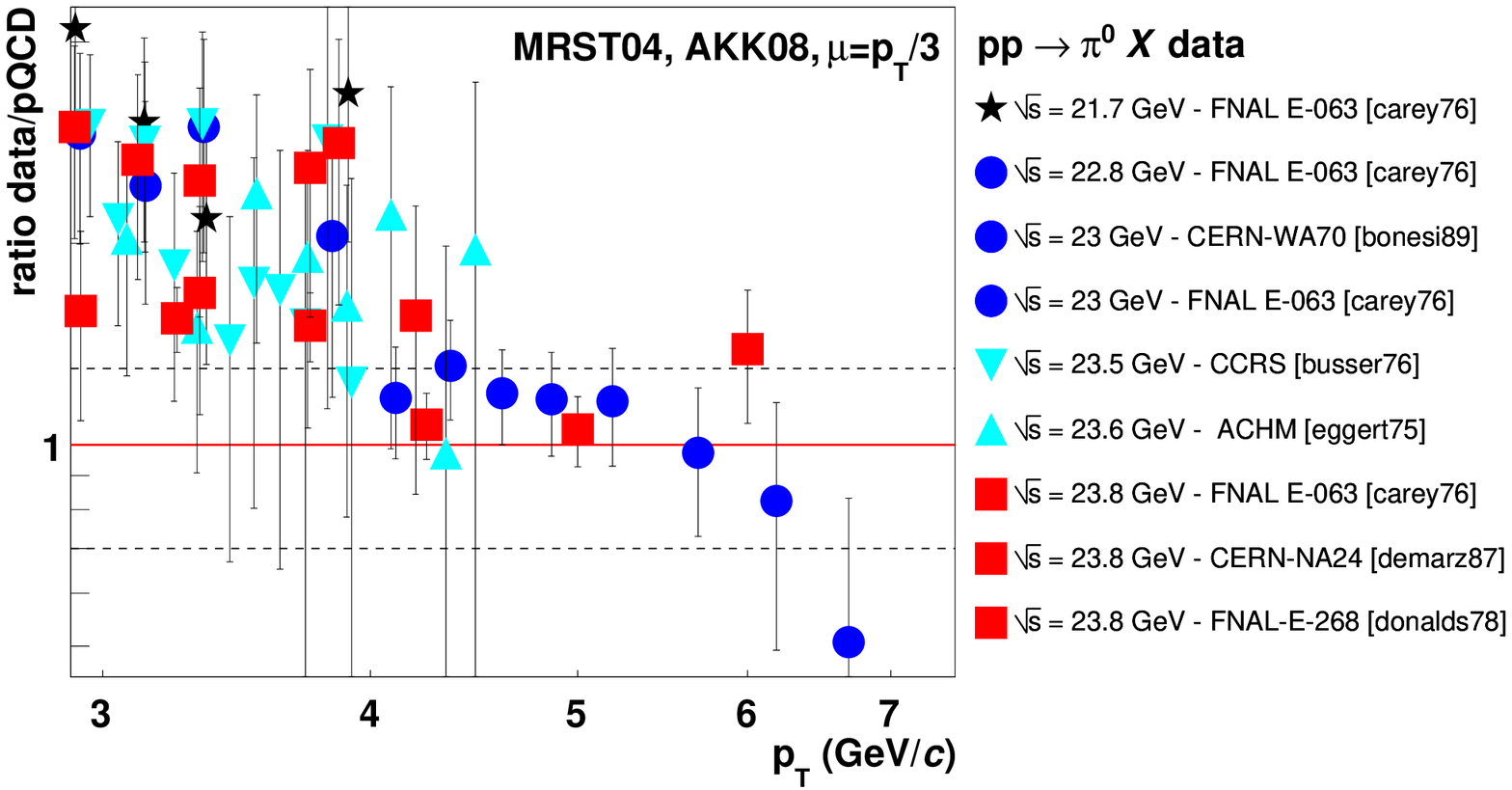}
\includegraphics[width=8.9cm,height=4.3cm]{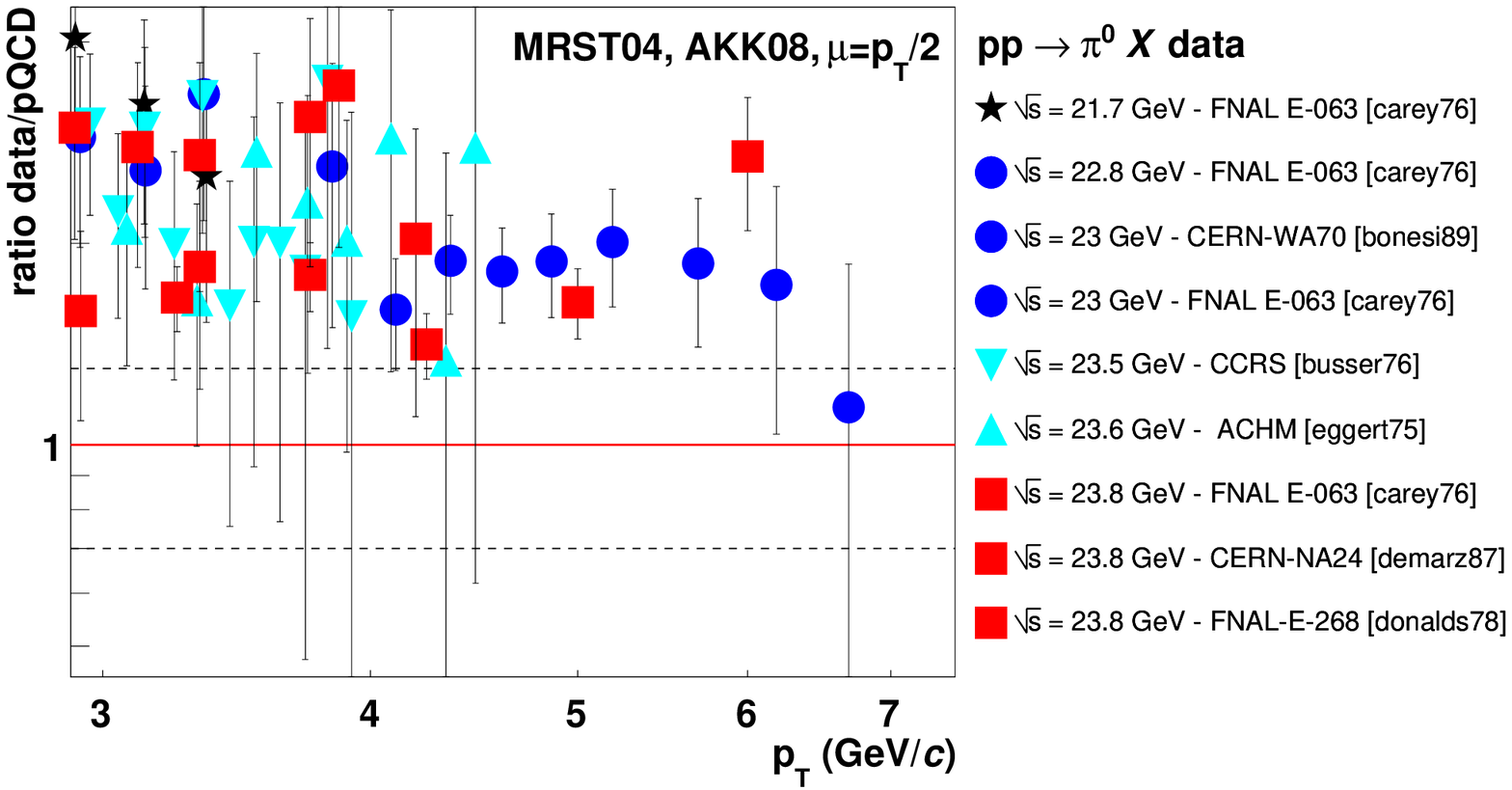}\\
\includegraphics[width=6.0cm,height=4.3cm,bb=0 0 360 304,clip]{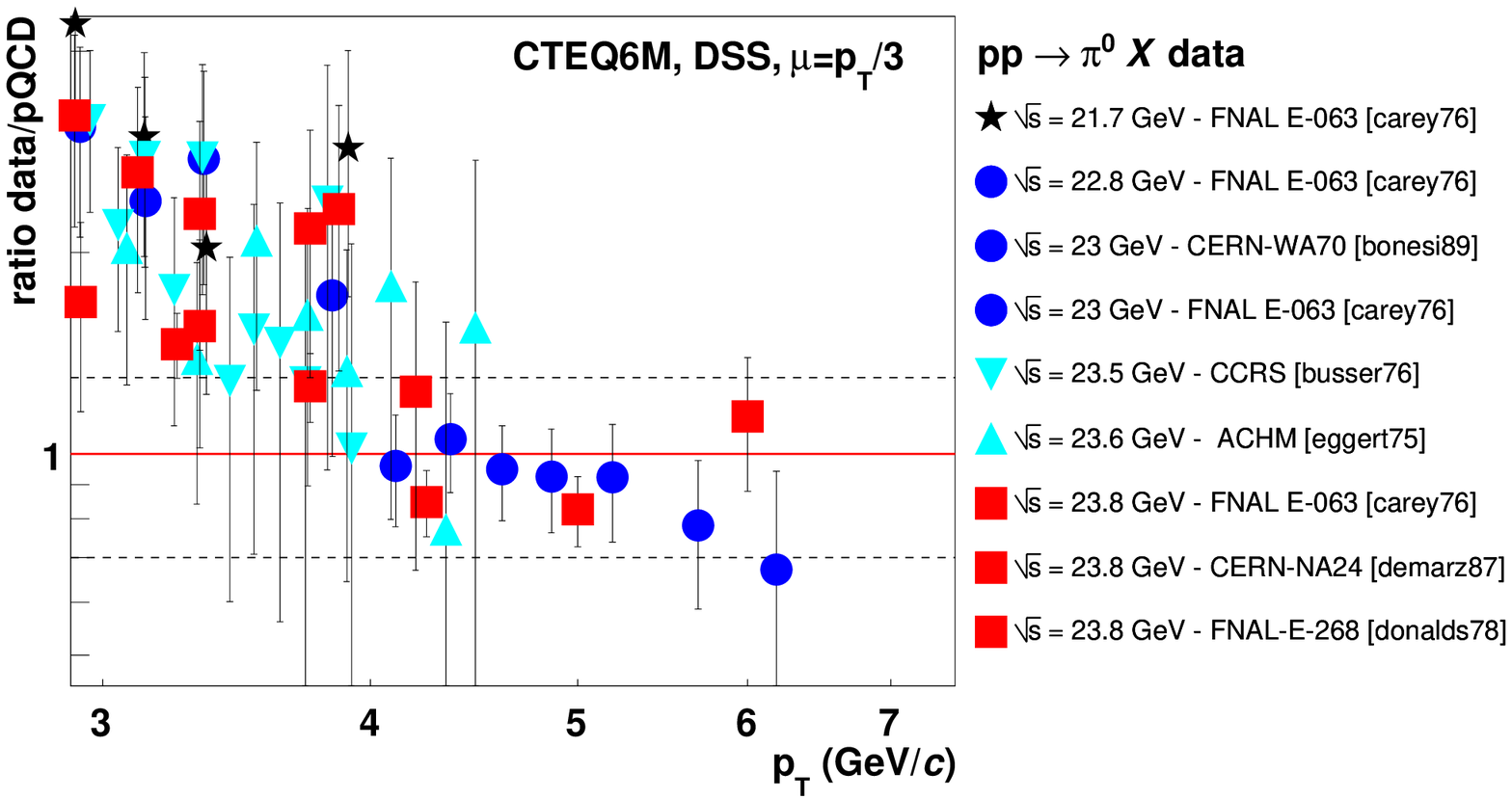}
\includegraphics[width=8.9cm,height=4.3cm]{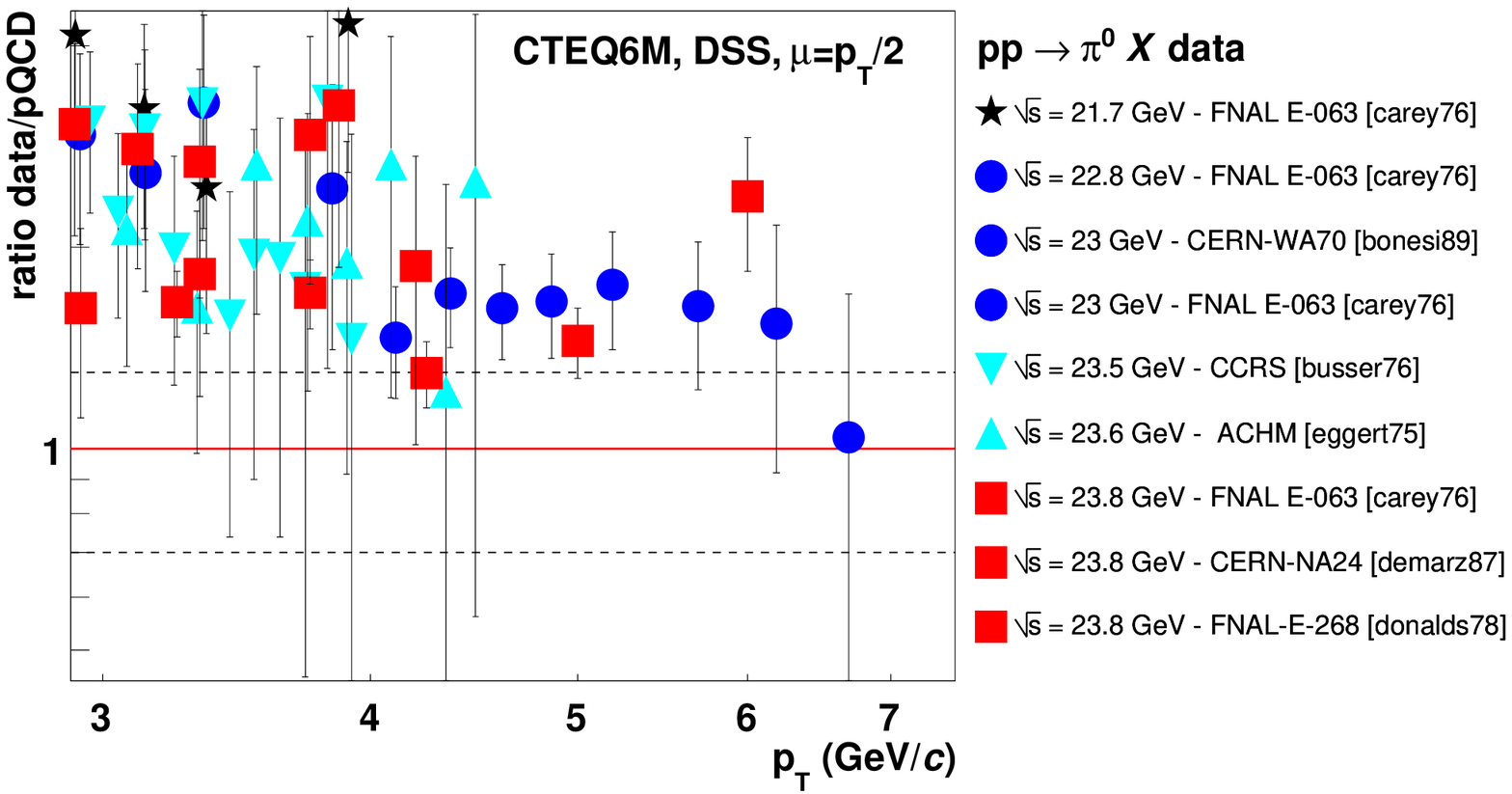}\\
\includegraphics[width=6.0cm,height=4.3cm,bb=0 0 360 304,clip]{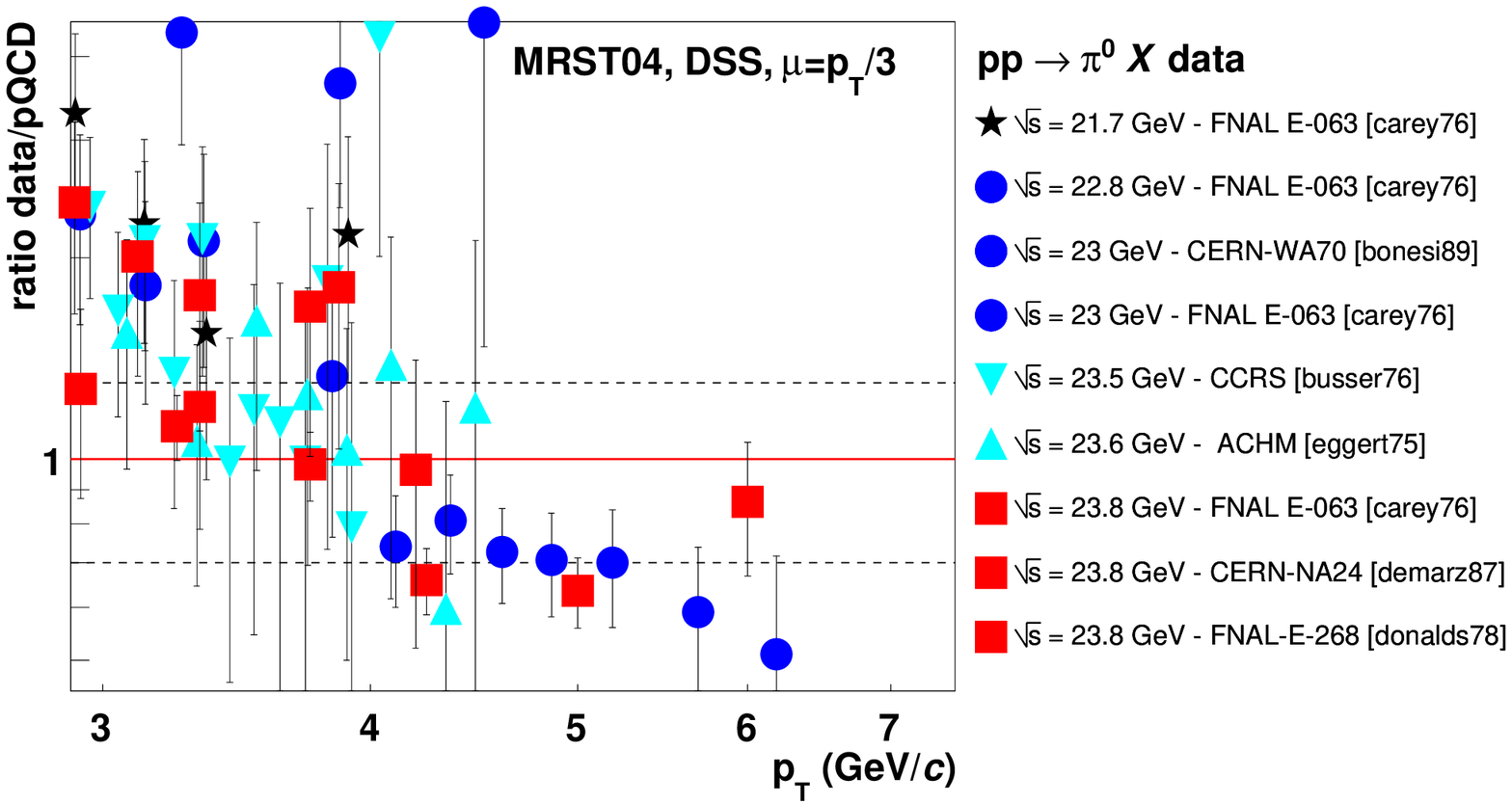}
\includegraphics[width=8.9cm,height=4.3cm]{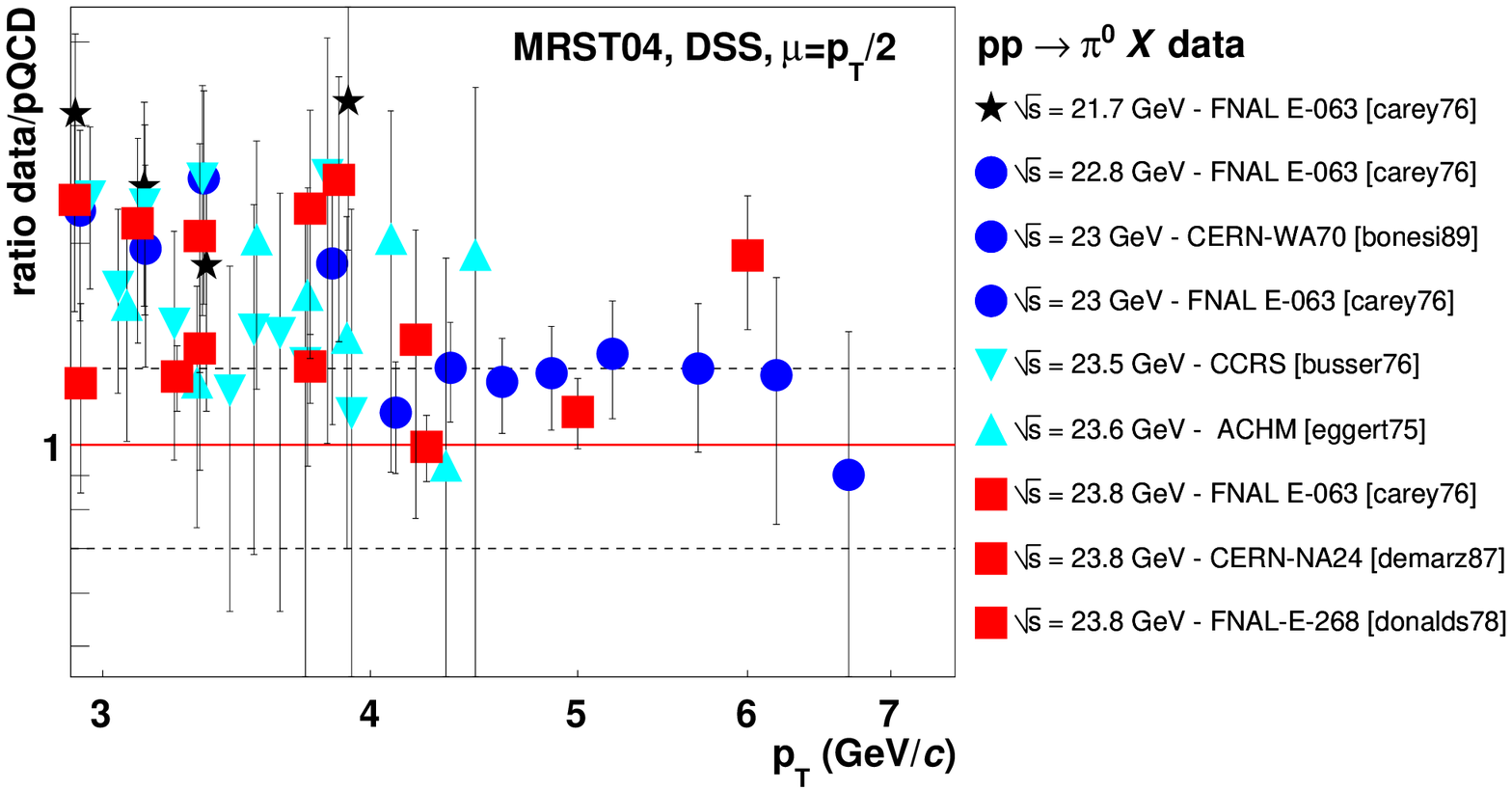}
\end{center}
\caption{Ratio of data over pQCD for pion transverse spectra in $p$--$p$ collisions at $\sqrt{s}\approx$~21.7--23.8~GeV. 
The left (right) plots are for theoretical scales $\mu~=~p_{_T}/3$ ($p_{_T}/2$). Two sets of PDFs 
(MRST04 and CTEQ6.1M) and three FFs (AKK05, AKK08, DSS, from top to bottom) are used.
The dashed lines are just indicative for variations of $\pm$30\% from the reference at R~=~1.}
\label{fig:ratio_data_pQCD}
\end{figure}

At large values of $p_{_T}$, the use of the fixed-order perturbation theory is fully justified, 
since the perturbative series is controlled by a small expansion parameter, $\alpha_s(p_{_T}^2)$.
However, in the typical kinematic range of fixed-target experiments, where
$x_{_T}\equiv2p_{_T}/\sqrt{s}\gtrsim$~0.1, the coefficients of the perturbative expansion are enhanced 
by extra powers of logarithmic terms of the form $\alpha_s^n\ln^{2n}(1-x_{_T})$ or $\alpha_s^n\ln^{2n-1}(1-x_{_T})$. Resummation to all orders of 
such ``threshold'' terms -- which appear because the initial partons have just enough energy to 
produce the high-transverse momentum parton -- have been carried out
at next-to-leading logarithmic (NLL) accuracy in~\cite{deFlorian:2005yj,deFlorian:2007ty}. 
These studies confirm that accounting for these terms results in a large (approximately $p_{_T}$-independent) 
enhancement of the perturbative cross section for pion production in the range of fixed-target 
energies of relevance here ($\sqrt{s}\approx20$~GeV). These studies also find that the scale 
dependence is also reduced at NLL compared to NLO. The presently used fixed-order calculations (INCNLO) do not include 
threshold resummations but their effect in the final spectrum is accounted for, in an effective way,
by our choice of relatively small theoretical scales, $\mu/p_{_T}=1/2$--$1/3$, which results
in a cross section increase of a factor of $\sim$2--3 as compared to the $\mu/p_{_T}=1/2$--$2$ range 
used e.g. in~\cite{deFlorian:2005yj,deFlorian:2007ty}.\\

The two non-perturbative inputs of Eq.~(\ref{eq:dsigma_pQCD}) are the parton 
densities and the fragmentation functions. The former are obtained mainly 
from global-fit analyses of deep-inelastic electron-proton data, the latter from hadron 
production results in $e^+e^-$ collisions. The PDFs are known to within $\sim$20\% uncertainty~\cite{CTEQ6.1M}
in the kinematic range of interest here: $x_{_T}=p_{_T}/p_{_T}^{\rm max}\approx$~0.2--0.5 at midrapidity. 
We use here two of the latest standard PDFs available: MRST04~\cite{mrst04} and 
CTEQ6.1M~\cite{CTEQ6.1M}. 
For  the quark and gluon fragmentation functions into pions, we use and compare three parametrizations: the commonly used AKK05~\cite{akk} plus two more recent sets:
DSS~\cite{dss} and AKK08~\cite{akk08}. 
The dominant fragmentation contribution to Eq.~(\ref{eq:dsigma_pQCD}) comes from the 
large-$z$ domain: $\mean{z}=\mean{p_{\rm hadron}/p_{\rm parton}}\approx$~0.8 
for $p_{_T}\gtrsim$~3~GeV$/c$ at $\sqrt{s}~=~22.4$~GeV,
where the $e^+ e^-$ fragmentation data used to obtain the FFs are scarce. 
In addition, the gluon-to-pion FF is not well determined by $e^+e^-$ annihilation data, 
as it appears there only at NLO, and we explore small fragmentation scales (in particular, 
when using $\mu_{\,\ensuremath{\it{ff}}}=p_{_T}/3$) far away from the kinematical regions where the $e^+e^-$ 
fits are performed. All these issues, which were a concern for the older FF sets like 
KKP~\cite{kkp}, Kre~\cite{kretzer} or AKK05, have been partially 
solved with the most recent fits~\cite{albino:2008af} which include for the first time also
hadronic data (and error analyses, such as for HKNS~\cite{hkns}) in their global analyses. These new fits cover a larger $z$ range and are 
more sensitive to the gluon fragmentation. As a result, the normalization of the gluon 
fragmentation function into pions is increased by e.g. up to 50\% in AKK08~\cite{akk08} with respect to AKK05~\cite{akk} at the $Z^0$ mass scale. This has an obvious 
impact in the absolute normalization of the predicted pion spectra as we discuss below.\\

In Figs.~\ref{fig:data_pQCD} (spectra) and \ref{fig:ratio_data_pQCD} (ratio data/pQCD), 
the measured pion $p$--$p$ single inclusive distributions at various energies are compared to the corresponding 
NLO predictions for varying theoretical scales ($\mu=p_{_T}/3$ and $p_{_T}/2$),  PDFs (MRST04 and CTEQ6.1M) 
and FFs (AKK05, AKK08 and DSS). In general, the calculations tend to underpredict the measured
cross sections. The overall agreement, in the $p_{_T}$ dependence and absolute normalization, improves 
going from the left (scales $\mu=p_{_T}/3$) to the right (scales $\mu=p_{_T}/2$) and when using 
MRST instead of CTEQ. The MRST04 parametrization results in a cross section 25\% 
larger than using CTEQ6.1M in the range\footnote{However, closer to the kinematical limit, above 8~GeV$/c$, 
the trend changes rapidly and the CTEQ6.1M fit overshoots the MRST04 one by up to 40\%, indicating 
the large current uncertainty of the gluon and sea-quark densities at high values of $x$.} $p_{_T}$~=~3--6~GeV$/c$. 
Such a difference in the resulting cross sections is larger than expected from error analysis within 
a single PDF set. The AKK08 and DSS fragmentation functions reproduce better the data than the 
AKK05 ones. The overall trend is consistent with MRST04 and AKK08/DSS predicting a {\it higher} 
pion yield than CTEQ6.1 and AKK05 in the kinematic range of interest here.
In any case, the data-theory agreement at fixed-target energies for high-$p_{_T}$ pions is clearly 
better than for prompt-photon production, where the measured E706 yield at $\sqrts =31.6$--$38.6$~GeV 
appears to be two to three times larger than the corresponding INCNLO predictions~\cite{Aurenche:2006vj}. 


\section{Inclusive pion spectra in $p$--$p$ collisions at $\sqrt{s}$~=~22.4~GeV: a practical parametrization}
\label{sec:fit}

After verifying that the fixed-order pQCD calculations can reproduce relatively well the existing high-$p_{_T}$
pion data at fixed-target energies, the second motivation of this study is to provide a practical parametrization 
of the $p$--$p$ pion spectrum at $\sqrt{s}$~=~22.4~GeV to be used as reference baseline for high $p_{_T}$ 
$\pi^0$ production in A-A collisions at the same c.m. energy, where no proton-proton data 
has been yet measured at RHIC~\cite{Adare:2008cx}. We discuss here the method followed to obtain 
a fit from the existing experimental data sets after rescaling them to a common center-of-mass energy 
making use of the NLO predictions.

\subsection{Center-of-mass energy rescaling}
\label{sec:rescaling}

\begin{figure}[htbp]
\begin{center}
\includegraphics[height=7.5cm,width=8.cm]{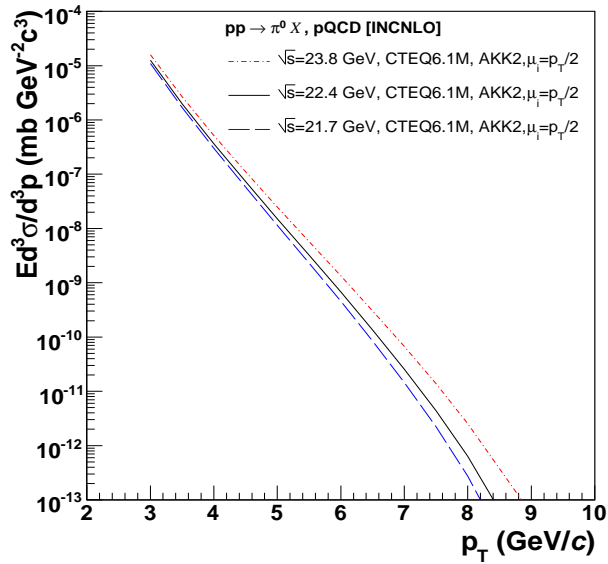}\hfill
\end{center}
\vspace{-0.5cm}
\caption{Differential $\pi^0$ cross sections in $p$--$p$ collisions predicted by NLO pQCD 
calculations with scales $\mu = p_{_T}/2$, CTEQ6.1M parton-distribution-functions, and AKK08
fragmentation functions
at $\sqrt{s}$~=~21.7, 22.4 and 23.8~GeV.}
\label{fig:nlo_21_23}
\end{figure}

The existing data sets (Table~\ref{tab:compilation}) cover the range of c.m. energies from 21.7~GeV to 23.8~GeV. 
Although at low-$p_{_T}$ (below $\sim$~2~GeV$/c$), the small differences in $\sqrts$ result into negligible 
variations of the soft pion yield and all spectra agree well (see Fig.~\ref{fig:all_spectra}), at high $p_{_T}$ 
-- as one approaches the kinematical limit -- a couple of GeV of extra c.m. energy available can result into a significant 
change in the parton-parton cross sections. For instance, 
as can be seen in Fig.~\ref{fig:nlo_21_23}, at $p_{_T}$~=~5~GeV$/c$, going from $\sqrt{s}=22.4$~GeV 
up (down) to 23.8~GeV (21.8~GeV) results in an increase (decrease) of the cross section by a factor of 
$\sim 60\%$ ($\sim 30\%$).

\begin{figure}[ht]
\begin{center}
\includegraphics[height=7.0cm,width=10.cm]{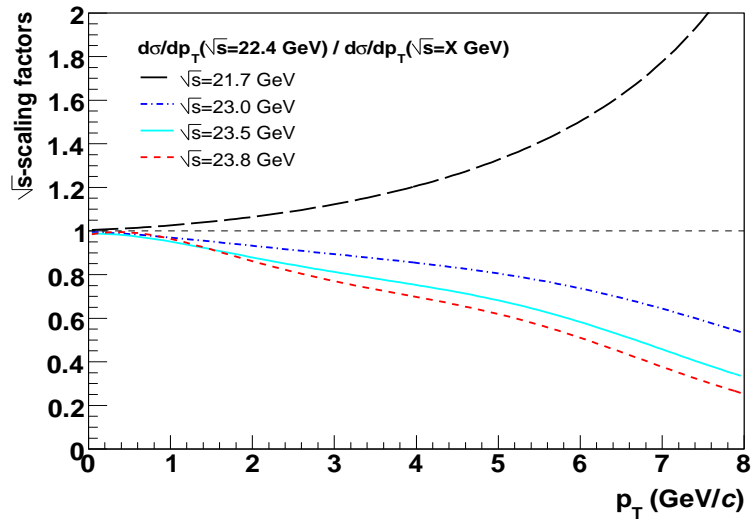}\hfill
\end{center}
\vspace{-0.5cm}
\caption{Rescaling correction factors of the pion cross-sections at c.m. energies 21.7--23.8~GeV to
a common $\sqrt{s}=22.4$~GeV value, as a function of $p_{_T}$, obtained from the ratio of NLO 
calculations given by Eq.~(\protect\ref{eq:rescaling}). }
\label{fig:rescale_comp}
\end{figure}

Although, as seen in the previous section, there are relatively large uncertainties in the NLO 
predictions for the {\it absolute} cross sections,  
most of these uncertainties cancel out when taking {\it ratios} of the predicted perturbative yields 
at different, yet close, c.m. energies. In particular, the (large) scale dependence is completely 
removed. One can, thus, rescale all experimental data points measured 
at a given $\sqrt{s}$~=~$X$~GeV (in the range 21.8 -- 23.8~GeV) to a common 
$\sqrt{s}$~=~22.4~GeV value via the following prescription:
\begin{equation}
\frac{\dd\sigma_{\rm exp}~(\sqrt{s}=22.4~{\rm GeV})}{\dd p_{_T}} = 
\left(\frac{\dd\sigma_{_{\ensuremath{\it{{\rm NLO}}}}}/\dd p_{_T}~(\sqrt{s}=22.4~{\rm GeV})}{\dd\sigma_{_{\ensuremath{\it{{\rm NLO}}}}}/\dd p_{_T}~(\sqrt{s}=X~{\rm GeV)}}\right)
\times \frac{\dd\sigma_{\rm exp}~(\sqrt{s}=X~{\rm GeV})}{\dd p_{_T}}.
\label{eq:rescaling}
\end{equation}
The pQCD cross-sections are computed in the range $p_{_T}\approx3$--$10$~GeV$/c$ for the 
4 energies under consideration and the ratio over the predictions at $\sqrts=22.4$~GeV 
is fitted to a polynomial form of order 2 or 4. Obviously, to minimize the theoretical uncertainties, 
both the denominator and numerator of the NLO ``rescaling factor'' (the expression in parentheses 
in Eq.~(\ref{eq:rescaling})) need to be computed using consistently the same PDFs, FFs and scales. 
The scaling factors provided here are obtained averaging over various different choices of these ingredients. 
The resulting scaling factors differ, in any case, as expected by a very small factor $\pm$5\%, 
well covered within the experimental uncertainties alone. The functional form of the rescaling 
factor is chosen so that the correction is zero at $p_{_T}$~=~0~GeV$/c$, 
so as to obtain a smooth extrapolation in the low-$p_{_T}$ region. In any case, below $p_{_T}\approx$ 1~GeV$/c$, 
the correction is (well) below $\sim$5\% and, so, the experimental low-$p_{_T}$ points are virtually 
unmodified as they should be by applying this rescaling procedure. The final correction functions are 
shown as a function of $p_{_T}$ in Fig.~\ref{fig:rescale_comp}.

In order to better estimate the uncertainty of the rescaling factors computed theoretically, the energy rescaling 
has also been determined  {\it a posteriori}, assuming that the invariant production cross section is a scaling 
function of $x_{_T}\equiv2p_{_T}/\sqrt{s}$:
\begin{equation}\label{eq:xt_scaling}
E\ \frac{\dd^3\sigma(pp\to\pi{\rm X})}{\dd^3 p}~(\sqrt{s})\ \propto \left(\frac{1}{\sqrt{s}}\right)^4\ F(x_{_T}),
\end{equation}
as it should be in perturbative QCD. Taking for the function $F$ the final parametrization discussed in the next 
Section, $F(x_{_T})=(22.4\ {\rm GeV})^4 f(x_{_T}\times [11.2\ {\rm GeV}])$, the rescaling factor is computed 
using Eq.~(\ref{eq:xt_scaling}). The difference between this empirical estimate and the theoretical rescaling factors, 
roughly $10\%$, is assigned as the uncertainty of the present energy rescaling procedure.

\subsection{Global fit of the rescaled pion $p_{_T}$ spectra at $\sqrt{s}=22.4$~GeV}
\label{sec:globalfit}

\begin{figure}[htbp]
\begin{center}
\includegraphics[height=10.0cm,width=10.cm]{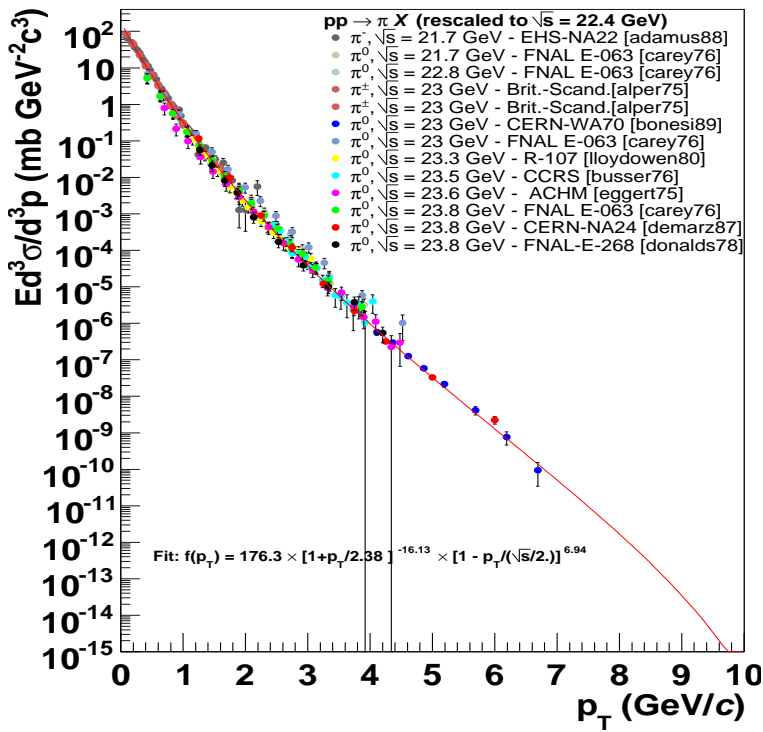}\end{center}
\caption{Compilation of all pion transverse spectra measured in $p$--$p$ collisions at 
$\sqrt{s}$~=~21.7 --~23.8 GeV, rescaled to a common $\sqrt{s}$~=~22.4~GeV energy, 
as discussed in the text, and fitted to Eq.~(\ref{eq:final_fit}), with the parameters~(\ref{eq:params}).}
\label{fig:final_fit}
\end{figure}

By applying the appropriate energy correction factors discussed in the previous section
to all the experimental data sets, we obtain a new set of data points which approximates better 
the expected $\pi^0$ spectrum at $\sqrt{s}=22.4$~GeV. The experimental spectrum 
$\frac{E \dd^3\sigma}{\dd^3p}\Bigr\vert_{y=0}$ 
is fitted to the following empirical 4-parameter functional form:
\begin{equation}   
f(p_{_T},\{p_i\}_{i=0,3})\ =\ p_0\cdot [ 1 + (p_{T}/p_1)]^{p_2}\cdot [ 1 - (p_{T}/p_{T}^{\rm max})]^{p_3} 
\label{eq:final_fit}
\end{equation}
Such a formula 
interpolates well between the low-$p_{_T}$ exponential shape and the high-$p_{_T}$ power-law 
while fulfilling the requirement of being zero at the kinematical limit ($p_{_T}^{\rm max}$~=~11.2~GeV/$c$, 
fixed in the fit). The $p_0$ parameter gives the cross-section at zero $p_{_T}$, $p_1$ 
indicates the transition value from soft to hard production, and the $p_2$ and $p_3$ exponents 
characterize the power-law and end of phase-space ranges.
After rejecting two data sets which are not consistent with the rest of spectra (see below), 
we obtain a final set of $n_{\rm dat}=194$~data points fitted with Eq.~(\ref{eq:final_fit}). 
The resulting fit is shown in Fig.~\ref{fig:final_fit}. The parameters are obtained from the minimization of the $\chi^2$ function
\begin{equation}
  \label{eq:chi2}
  \chi^2(\{p_i\}) = \sum_{j = 1}^{n_{\rm dat}} \ \left[\frac{\frac{E \dd^3\sigma}{\dd^3p}\Bigr\vert_{y=0}(p_{T_j}) - f(p_{T_j}, \{p_i\})}{\sigma_{j}}\right]^2,
\end{equation}
where $\sigma_j$ is the statistical and systematic error of point $j$ added in quadrature. The error of the parameters $p_i$ are given from a deviation of $\Delta\chi^2$ from its minimum: 
\begin{equation}
  \label{eq:dchi2}
  \chi^2(\{p_i+\delta p_i\}) - \chi^2(\{p_i\}) = \Delta\chi^2.
\end{equation}
Although a usual choice is $\Delta\chi^2=1$, we shall conservatively allow for a larger variation of the fit parameters assuming $\Delta\chi^2=50$ in what follows\footnote{This would correspond to an increase of $25\%$ of $\chi^2_{\rm min}/\ndf$ with $\ndf\simeq 200$.}, similarly to what is done in global fit analyses of parton densities or fragmentation functions (see e.g.~\cite{dss,Pumplin:2001ct}). From this procedure, the corresponding parameters are:
\begin{eqnarray}\label{eq:params}
p_0  &=& 176.3    \pm    69.7\;\;\mbox{[mb~GeV}^{-2}c^{3}]\nonumber\\
p_1 &=& 2.38\pm 1.19 \;\;\mbox{[GeV/$c$]}\nonumber\\
p_2  &=&-16.13\pm 7.21\\
p_3 &=& 6.94\pm 5.64\nonumber\\
\chi^2/\ndf&=&208.2/190\nonumber
\end{eqnarray}
with an important correlation between parameters and errors, as indicated by the large non-diagonal terms of the covariance (error) matrix, $V_{ij}$:
\begin{equation*}
V_{ij}= \begin{pmatrix}
          1.000 &-0.725 &-0.603 & 0.394\\
         -0.725 & 1.000  &0.981 &-0.862\\
         -0.603  &0.981 & 1.000& -0.940\\
          0.394& -0.862& -0.940  &1.000
\end{pmatrix}
\end{equation*}

Note that at low $p_{_T}$, this fit is consistent with an exponentially decreasing function with inverse slope 
$-p_1/p_2=148\pm 16$~MeV. The scale $p_1$, which naively separates soft from hard dynamics, has a sensible 
value $\sim 2$--$3$~GeV. Finally, the negative power slope, $p_2=-16.13$, is found to be larger in 
absolute value than $p_2=-10$ obtained at $\sqrt{s}=200$~GeV~\cite{phenix_pp_200GeV}. This is expected 
from the steeper dependence of parton densities and fragmentation functions probed at higher $x$ and $z$, 
respectively, at lower $\sqrt{s}$. The uncertainty $\Delta f$ of the fit is given by
\begin{equation}\label{eq:errprop}
\Delta f = \left[\ \sum_{i,j=0}^3\ \frac{\partial f}{\partial p_i}\ V_{ij}\ \frac{\partial f}{\partial p_j}\ \right]^{1/2}.
\end{equation}
The relative uncertainty of the parametrization, $\Delta f /f$, spans the range from $\sim$15\% at low $p_{_T}\lesssim 2$~GeV$/c$ up to 25\% (40\%) at $p_{_T}=4$~GeV$/c$ (5~GeV$/c$) in the range covered by the RHIC measurements~\cite{Adare:2008cx}.
At higher $p_{_T}$'s the fit is completely unconstrained due to the lack of data, and its uncertainty is very large.

\begin{figure}[htpb]
\begin{center}
\includegraphics[height=9.0cm]{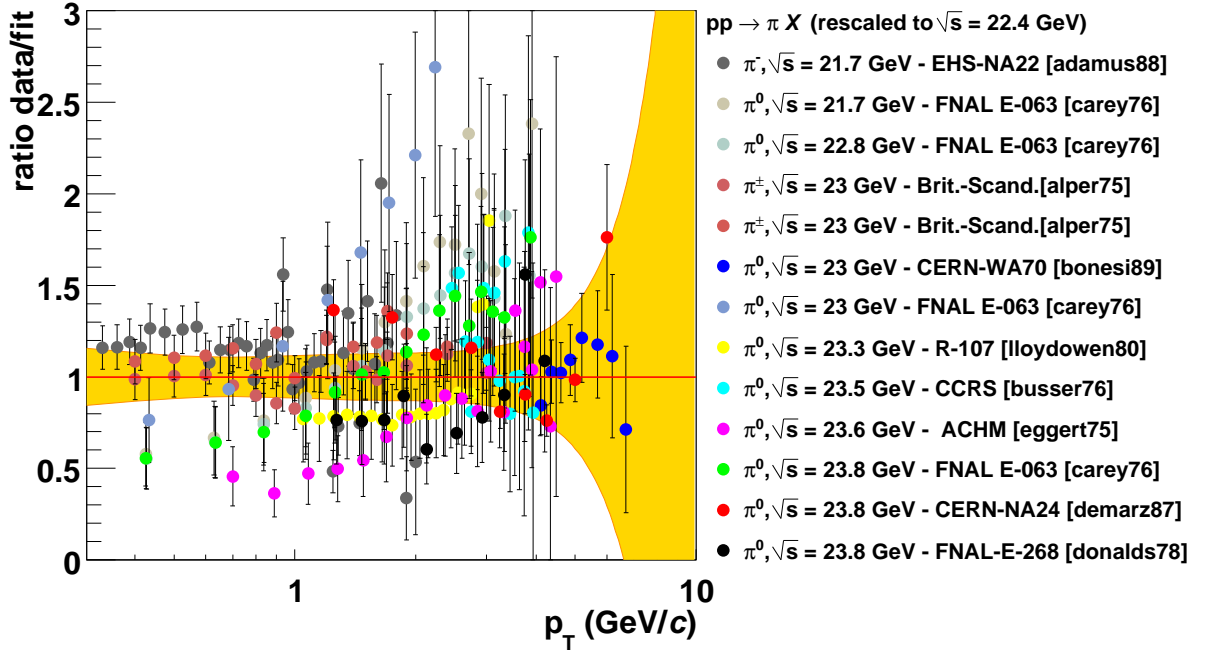}\hfill \end{center}
\caption{Ratio of all pion transverse spectra measured in $p$--$p$ collisions rescaled to
$\sqrt{s}$~=~22.4~GeV over the fit given by Eq.~(\protect\ref{eq:final_fit})
with the parameters~(\protect\ref{eq:params}). The yellow band 
represents the uncertainty assigned to the parametrization, as given by Eq.~(\ref{eq:errprop}). 
The [carey76] and [eggert75] points at $\sqrt{s}=23.6, 23.$~GeV respectively~\cite{carey76,eggert75}
have been excluded from the global fit.}
\label{fig:ratio_data_over_fit}
\end{figure}

Figure~\ref{fig:ratio_data_over_fit} shows the ratio of all data sets compiled and rescaled 
in this work over the fit Eq.~(\ref{eq:final_fit}) with the parameters quoted above. All data sets -- 
but Carey76~\cite{carey76} at $\sqrts$~=~23 GeV and Eggert75~\cite{eggert75} at $\sqrts$~=~23.6 GeV
which have a shape and absolute normalization inconsistent with the rest of measurements and have 
not been included in the final global analysis -- show a rather good 
agreement with the proposed parametrization, as also indicated by  $\chi^2/\ndf\simeq 1.1$.
We note that our empirical fit includes 
also the low-$p_{_T}$ range, not amenable to perturbative analysis, since we want to provide 
a (potentially useful) $p$--$p$ reference parametrization in the whole range covered by the 
nucleus-nucleus data.

\section{Summary}

We have compared the available high-$p_{_T}$ pion spectra measured in proton-proton collisions 
in the range  $\sqrt{s}$~=~21.7 --~23.8~GeV (CERN-ISR collider and CERN and FNAL fixed-target) 
to next-to-leading order pQCD calculations with recent parton distribution functions (PDFs) 
and fragmentation functions (FFs). A choice of the theoretical (factorization, fragmentation and 
normalization) scales between $p_{_T}/3$ and $p_{_T}/2$ reproduces well the magnitude and shape of the 
experimental data. CTEQ6.1 and MRST04 parton densities yield results different by up to 25\%. 
Second-generation parton-to-pion fragmentation functions (FFs) with updated constraints 
on the gluon and large-$z$ fragmentation region such as DSS or AKK08, improve the agreement of the data 
with the calculations compared to older FF parametrizations.\\

A baseline nucleon-nucleon reference $p_{_T}$-distribution for inclusive $\pi^0$ 
production at $\sqrt{s}=22.4$~GeV has been determined from a global fit
analysis of the available data. 
The measured (high-$p_{_T}$) data sets have been rescaled at a common c.m. energy 
making use of the predicted NLO pQCD yields at the various $\sqrt{s}$. The resulting 
parametrization is consistent within $\pm$15\% and $\pm$40\% systematic uncertainty with the 
rescaled $\pi^0$ and $\pi^\pm$ measurements at low ($p_{_T}\lesssim 2$~GeV/$c$) and 
moderate ($p_{_T}\simeq5$~GeV$/c$) transverse momentum.
Such a reference -- Eq.~(\ref{eq:final_fit}) with fit parameters~(\ref{eq:params}) --
can be used in order to obtain the nuclear modification factor of high-$p_{_T}$ pion 
production in A-A collisions at $\sqrt{s_{_{NN}}}$~=~22.4~GeV measured at RHIC.

\acknowledgments

\noindent
Valuable comments and discussions with Klaus Reygers and Mike J. Tannenbaum
are acknowledged. We thank Simon Albino for providing us with the latest AKK08
fragmentation functions. DdE is supported by the 6th EU Framework Programme 
(contract MEIF-CT-2005-025073). FA thanks the hospitality of CERN PH-TH department where part of this work has been completed.


\end{document}